\documentclass[12pt]{article}

\newenvironment{wileykeywords}{\textsf{Keywords:}\hspace{\stretch{1}}}{\hspace{\stretch{1}}\rule{1ex}{1ex}}

\usepackage{amsmath,amssymb}
\usepackage{graphicx}
\usepackage{color}
\usepackage{dcolumn}
\usepackage{bm}
\usepackage[numbers,comma,sort&compress]{natbib}
\usepackage{natmove} 

\usepackage{mathtools}
\usepackage{amsthm}
\usepackage{algorithmic,algorithm}

\usepackage{subcaption}
\usepackage{authblk}

\DeclareMathOperator*{\argmin}{\arg\!\min} 
\newcommand*{\argminl}{\argmin\limits} 

\def\X{{\mathcal{X}}}
\def\I{{\mathcal{I}}}
\def\J{{\mathcal{J}}}
\def\L{{\mathcal{L}}}

\def\PP{{\mathbb{P}}}

\def\tc{{\tilde{c}}}

\definecolor{background-color}{gray}{0.98}

\title{Bayesian Uncertainty Quantification in Inverse Modelling of Electrochemical Systems}

\begin{document}
\author[1]{Athinthra Sethurajan}
\author[2]{Sergey Krachkovskiy} 
\author[2]{Gillian Goward}
\author[3]{Bartosz Protas\thanks{Corresponding Author, Email: {\tt bprotas@mcmaster.ca}}$^{,}$} 
\affil[1]{School of Computational Science and Engineering, McMaster University, Hamilton, ON, Canada L8S 4K1}
\affil[2]{Department of Chemistry and Chemical Biology, McMaster University,  Hamilton, ON, Canada L8S 4K1}
\affil[3]{Department of Mathematics and Statistics, McMaster University, Hamilton, ON, Canada L8S 4K1}
\date{\today}

\maketitle

\begin{abstract}
  This study proposes a novel approach to quantifying uncertainties of
  constitutive relations inferred from noisy experimental data using
  inverse modelling. We focus on electrochemical systems in which
  charged species (e.g., Lithium ions) are transported in electrolyte
  solutions under an applied current. Such systems are typically
  described by the Planck-Nernst equation in which the unknown
  material properties are the diffusion coefficient and the
  transference number assumed constant or concentration-dependent.
  These material properties can be optimally reconstructed from time-
  and space-resolved concentration profiles measured during
  experiments using the Magnetic Resonance Imaging (MRI) technique.
  However, since the measurement data is usually noisy, it is
  important to quantify how the presence of noise affects the
  uncertainty of the reconstructed material properties. We address
  this problem by developing a state-of-the-art Bayesian approach to
  uncertainty quantification in which the reconstructed material
  properties are recast in terms of probability distributions,
  allowing us to rigorously determine suitable confidence intervals.
  The proposed approach is first thoroughly validated using
  ``manufactured'' data exhibiting the expected behavior as the
  magnitude of noise is varied. Then, this approach is applied to
  quantify the uncertainty of the diffusion coefficient and the
  transference number reconstructed from experimental data revealing
  interesting insights.
\end{abstract}

\begin{wileykeywords}
  electrolytic transport, Planck-Nernst equation, inverse problems, 
  Bayesian uncertainty quantification, Markov chain Monte-Carlo
\end{wileykeywords}

\section{\sffamily \Large INTRODUCTION} 



\label{sec:intro}

In this study we develop and validate a probabilistic framework for
quantifying uncertainty in the reconstruction of unknown material
properties of electrochemical system from experimental data using
inverse modelling.  Electrochemical systems have been studied for a
long time and play a major role in advancement of technology and the
way humans live. For instance, novel energy-storage solutions based on
Lithium-ion batteries have already revolutionized personal-electronics
industry \cite{broussely2004li} and are changing the automotive
industry \cite{lu2013review}. Lately this progress has increasingly
relied on mathematical models of the transport of charged species
which are typically derived from the Planck-Nernst equation
\cite{nt04}. These models crucially depend on a number of material
properties such as, e.g., the diffusion coefficients of active ions in
electrodes and electrolyte, specific to the different materials used
in electrochemical systems. The material properties of interest to us
here represent the {\em constitutive relations} describing how
thermodynamic fluxes depend on the corresponding thermodynamic forces.
Unfortunately, for many materials, especially new ones, they are
notoriously difficult to obtain either from first principles or via
direct measurements, which hampers the modelling efforts.

One remedy to this situation is offered by the methods of {\em inverse
  modelling} which integrate in a systematic matter measurements of a
system with its mathematical model in order to infer certain unknown
properties of the system. While inverse modelling has seen many
successful applications in natural sciences (see, e.g., the monograph
\cite{t05} for some examples), its applications in the field of
electrochemistry have been rather limited and as some notable
exceptions we can mention the studies \cite{yu1999determination,
  prosini2002determination,klett2012quantifying,rahman2016electrochemical}.
From the mathematical point of view, inverse problems are often
formulated in the optimization setting where a suitable error
functional representing the mismatch between the actual measurements
and the observations predicted by the model is minimized with respect
to the unknown material property \cite{ehn96,v02}. This is the
approach we followed in our earlier study
\cite{sethurajan2015accurate} in which we developed and validated an
inverse-modelling approach allowing one to extract the effective
diffusivity and the transference number characterizing an electrolyte
solution from space- and time-resolved measurements of the
concentrations of the charged species in a galvanostatic experiment. A
key novelty of this approach as compared to earlier studies is that
the material properties are inferred in a very general continuous
setting subject only to minimal assumptions. This problem can be
therefore viewed as learning an optimal form of a nonlinear
constitutive relation from data

Since the measurements are usually contaminated with noise, an
important question is how this affects the accuracy of the
reconstructed material properties. The reason is that inverse problem
often tend to be ``ill-posed'' \cite{ehn96}, meaning that small
modifications of the input data (measurements) may result in
significant changes of the obtained solution (here, the reconstructed
material properties). Therefore, in order to have confidence in the
obtained results, it is necessary to quantify how the measurement
uncertainty translates into the uncertainty of the reconstructed
material properties and, if more than one quantity is reconstructed
(as was the case in \cite{sethurajan2015accurate}), whether the
uncertainties of the reconstructed quantities are mutually correlated.
An emerging approach which casts the problem of uncertainty
quantification in probabilistic terms is Bayesian inference. In this
framework, which blends prior hypotheses on unknown parameters with
information from measurements in a systematic manner, the
reconstructions of parameters are given in terms of suitable
probability densities. General references to Bayesian inference
include \cite{s13,t17}, whereas a more general perspective which also
involves continuous problems described by differential equations was
developed in \cite{s10}. Recently, there has been a growing interest
in Bayesian approaches to the solution of inverse problems with
applications in electrical impedance tomography
\cite{kaipio2000statistical}, atmospheric science
\cite{bergamaschi2000inverse, bousquet1999inverse}, contaminant source
identification \cite{michalak2003method}, ground water modelling
\cite{rubin2010bayesian}, etc. However, in the field of
electrochemistry such techniques are not very common and have been
applied to quantify uncertainty in diagnostics and prognostics of
batteries~\cite{saha2009prognostics} and state estimation in battery
management systems~\cite{samadi2013online}. The goal of the present
investigation is to develop and validate a Bayesian approach to
uncertainty quantification in inverse reconstruction of
state-dependent material properties in electrochemical systems.  The
proposed approach is quite general and as such may be applicable to a
broad range of similar problems in chemistry governed by macroscopic
models. The main novelty, and at the same time the largest difficulty
which had to be overcome, is that the uncertainty needs to be
quantified for material properties reconstructed in the {\em
  continuous} setting.

The structure of the paper is as follows: in the next section we
describe the class of electrochemical systems of interest to us,
review their models and the inverse-modelling approach, and then
introduce the Bayesian formulation of uncertainty quantification; the
proposed approach is validated using synthetic data in Section
\ref{sec:validation}, whereas an application involving actual
experimental data is presented in Section \ref{sec:application};
conclusions and final remarks are deferred to Section \ref{sec:final}.

\section{\sffamily \Large METHODOLOGY}



\label{sec:method}

In this section we describe different constituents of our methodology:
we start with the measurement data, then introduce the Planck-Nernst
system as a mathematical model for the problem, after that we review
the inverse-modelling approach which is followed by the presentation
of a Bayesian strategy for uncertainty quantification.

\subsection{\sffamily \Large Experimental Measurements}
\label{sec:experiment}

\begin{figure}
\centering
\includegraphics[width=1\textwidth]{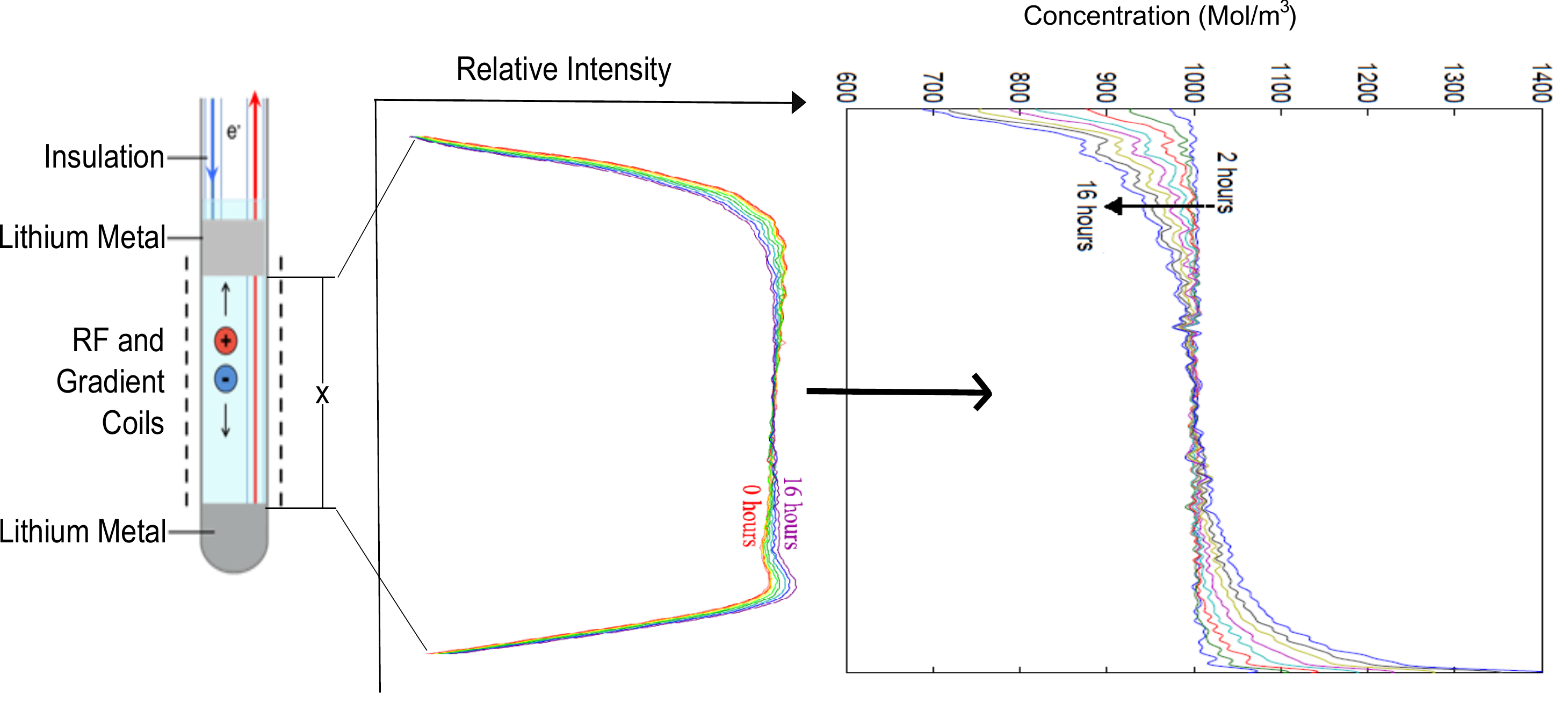}
\caption{Experimental setup and the measured concentration data
  $\tc(x,t)$ used in the present study.}
\label{fig:exp}
\end{figure}

To fix attention, we focus on a galvanostatic experiment in an
electrochemical cell for which the set-up is shown schematically in
Figure \ref{fig:exp}. The experiment monitors the gradual build-up of
the ionic concentration gradient in an electrolyte solution which
results from the application of a constant current, starting from an
initially uniform concentration throughout the solution volume. The
experiment is carried out under galvanostatic conditions in a
symmetric Li-Li electrochemical cell constructed from a 17 mm long and
4.2 mm diameter NMR tube, shown in Figure \ref{fig:exp}, filled with a
1 M LiTFSI solution in Propylene Carbonate (PC). A constant current of
50 $\mu$A was applied to the cell for 16 hours. Concentration profiles
were acquired using magnetic resonance imaging (MRI). For this
experiment we chose to monitor the $^{19}$F nuclei, which
significantly reduces the data acquisition time in comparison to
monitoring the $^7$Li nuclei, since the relative NMR sensitivity to a
$^{19}$F nucleus is approximately 3 times higher than to a $^{7}$Li
nucleus.  One-dimensional $^{19}$F NMR images were obtained using a
gradient spin-echo pulse sequence with the magnetic field gradient
applied along the axis of the cell, with a 3 ms echo time and a 20
G/cm reading gradient. In the course of the experiment 256
frequency-domain points were collected over the spectral width of 200
kHz. The combination of the magnetic field gradient and spectral
resolution yielded a spatial resolution of 40 $\mu$m. A total of 64
scans with a relaxation delay of 3.5 s were collected for each image,
resulting in an acquisition time of 4 minutes per image. The imaging
measurement sequence was repeated at 2-hour intervals uniformly spread
over 16 hours duration of the galvanostatic experiment. The
experimentally obtained concentration profiles, hereafter denoted
$\tilde{c}(x,t)$, are shown in Figure \ref{fig:exp} at different times
$t \in [0, 16 \ \text{hours}]$ as functions of the space coordinate
$x$. Further details concerning this experiment can be found in
\cite{sethurajan2015accurate}.

\subsection{\sffamily \Large The Planck-Nernst Model}
\label{sec:PN}

Here we recall the classical Planck-Nernst model used to describe the
transport of charged species in dilute electrolytes \cite{nt04}. The
concentrations of cations and anions are denoted by $c_+$ and $c_-$
respectively. We make the following modelling assumptions in order to
obtain the mathematical description of the mass transport during the
galvanostatic experiment described in Section \ref{sec:experiment}
\cite{sethurajan2015accurate}:
\begin{itemize}
\item[A1:] isothermal conditions;
\item[A2:] the driving force for mass transport of a species is the
  gradient of its chemical potential;
\item[A3:] the lack of thermodynamic ideality (i.e., activity
  coefficient different from one) and the effect of the solution
  viscosity accounted for by an a priori undetermined dependence of
  the material properties on the salt concentration;
\item[A4:] ion transport occurs only in the axial direction and
  transport in the radial direction of the cell is negligible;
\item[A5:] the electrolyte solution is homogeneous at the beginning of
  the experiment;
\item[A6:] the system satisfies local
  electrical neutrality at every location in the bulk, which implies
  that $c_{+}=c_{-}=c$, were c is the salt concentration;
\item[A7:] mass transport occurs only by diffusion and migration in
  the applied electric field (i.e., convective transport is
  neglected);
\item[A8:] the cation flux at the two boundaries ($x$ = 0 and $x$ =
  $L$) corresponds to the applied electric current and results in
  lithium deposition and stripping, respectively \cite{nbl08,nt04}.
\end{itemize}

We therefore consider a 1D problem with the spatial coordinate $x \in
[0,L]$, where $L$ is the length of the electrolyte filled region in the cell,
and time $t \in [0,T]$, where $T$ denotes the duration of the
experiment. The above assumptions lead to the following partial
differential equation (PDE) describing mass transport in the
electrolyte solution \eqref{eq:ffull}, subject to the boundary conditions
\eqref{eq:fbc} and the initial condition \eqref{eq:fic}:
\begin{subequations}
\label{eq:PN}
\begin{alignat}{2} \label{eq:ffull}
\frac {\partial c}{\partial t}&=\frac {\partial }{\partial x} \left [{D}\frac{ \partial c}{\partial x}+ \frac {\left (1-{ t^{+}} \right )I}{FA}\right ] & \qquad & \text{in} \ (0,L) \times (0,T],\\
 \label{eq:fbc}
\left. \frac {\partial c}{\partial x} \right |_{x=0, L}&=- \frac {\left (1-{  t^{+}} \right )I} {{ D}FA} &&  \text{in} \ (0,T],\\
\label{eq:fic}
\left. c \right |_{t=0} &= c^{\text{init}} && \text{in} \ (0,L),
\end{alignat}
\end{subequations}
where $c^{\text{init}}$ is the initial concentration, $A$ is the
cross-sectional area of the cell, $F$ is Faraday's constant, whereas
$I$ denotes the applied constant current. We note that the effective
Fickian diffusion coefficient $D$ and the transference number $t^+$
are considered unknown and will be reconstructed from the experimental
data using the inverse modelling approach described in the following
subsection. In the standard Planck-Nernst theory, both the diffusion
coefficient $D$ and the transference number $t^+$ are assumed
constant. In addition to this formulation, we will also consider a
more general set-up with the diffusion coefficient and the transference
number depending on the concentration, i.e., $D=D(c)$ and
$t^+=t^+(c)$, which accounts for the thermodynamic non-ideality of the
electrolyte solution. To simplify our notation, we will denote the
pair of unknown material properties $m$, regardless of whether these
properties are constant ($m= [D,t^+]$), or concentration-dependent
($m= [D(c),t^+(c)]$). The solutions of system \eqref{eq:PN} then
define a map $\L$ from the material properties $m$ to the space- and
time-dependent concentrations, i.e.,
\begin{equation}
  c(x,t;m) = \L(m), \qquad 0 \le x \le L, \ 0 \le t \le T.
\label{eq:sys}
\end{equation}

\subsection{\sffamily \Large Inverse Modeling}
\label{sec:inverse}

The unknown material properties, $D$ and $t^+$, can be reconstructed
based on the assumed transport model \eqref{eq:PN} using the
concentration profiles obtained in the NMR experiment described in
Section \ref{sec:experiment}.  We will use the deterministic inverse
modelling approach developed and validated in
\cite{sethurajan2015accurate} in which the problem is framed as
minimization of an error functional representing the least-squares
deviation between the concentration values $c$ predicted by model
\eqref{eq:PN} for a given set of material properties $m$ and the
experimentally determined concentration values $\tilde c$. The error
functional can thus be represented as 
\begin{equation} 
\label{eq:J}
\mathcal{J}(m)=
\frac{1}{2}\sum_{i=0}^{N_T} \int_{0}^{L} \left[c(x,t_i;m)-\tilde{c}(x,t_i) \right]^{2}\,dx,
\end{equation}
where $N_T$ is the number of time levels $t_i$ where the concentration
profiles are acquired during the experiment. We will consider two
distinct formulations corresponding, respectively, to constant and to
concentration-dependent material properties.

When both $D$ and $t^+$ are constant, we obtain a simple unconstrained
optimization problem (which is exact in the limiting case of an ideal
solution, i.e., at very dilute salt concentrations)
\begin{equation*}
\text{P1}: \qquad [\widehat{D}, \widehat t^+] = \argminl_{[t^{+},D] \in \mathbb R^2}\mathcal{J}([D,t^+])
\end{equation*}
(henceforth carets ``$\widehat{\cdot}$'' will denote optimal
reconstructions).  Problem P1 can be solved in a straightforward
manner using commercially available software tools such as the
minimization routines in {\tt MATLAB}.  It was in fact already solved
in the seminal study by Klett et al.~\cite{klett2012quantifying} and is also
solved here as a preliminary step in a more complete analysis.

A more complicated optimization problem arises when both $D(c)$ and
$t^+(c)$ are concentration-dependent, which reflects the physics of
the problem in more realistic fashion,
\begin{equation*}
\text{P2}: \qquad [\widehat D(c),\widehat t^+(c)] = 
\argminl_{[t^{+}(c),\, D(c)] \in \mathcal X}\mathcal{J}\left([D(c),t^+(c)]\right),
\end{equation*}
where $\X$ denotes a suitable function space to which the pair $[D(c),
t^+(c)]$ belongs. The functions $D(c)$ and $t^+(c)$ are defined on the
interval $[c_{\alpha},c_{\beta}]$ bounded by some minimum and maximum
concentrations $c_{\alpha}$ and $c_{\beta}$, respectively.

We emphasize that, apart from smoothness and the behavior at the
endpoints (i.e., for $c \rightarrow c_{\alpha},c_{\beta}$), no other a
priori assumptions are made about the functional forms of $D(c)$ and
$t^+(c)$. In contrast to the simplified case (problem P1), the
computational approach required to solve the more realistic case
(problem P2) with concentration-dependent material properties is more
involved and necessitates specialized tools. This approach has the
general form of iterative gradient-based minimization
\begin{subequations}
\label{eq:desc}
\begin{alignat}{2}
D^{(n+1)}(c) &= D^{(n)}(c)-\xi_{D}^{(n)}\, \nabla_{D}\mathcal{J}\left(D^{(n)}(c),t^{+ (n)}(c)\right) &\qquad & n=1,2,\dots, 
\label{eq:descD} \\
t^{+(n+1)}(c) &= t^{+(n)}(c)-\xi_{t^{+}}^{(n)} \, \nabla_{t^{+}}\mathcal{J}\left(D^{(n+1)}(c),t^{+(n)}(c)\right) && n=1,2,\dots,
\label{eq:desct} \\
[D^{(1)}(c),t^{+(1)}(c)] &= [\widehat D,\widehat t^+], && \label{eq:descIG}
\end{alignat}
\end{subequations}
where $\nabla_{D}\mathcal{J}$ and $\nabla_{t^+}\mathcal{J}$ are the
gradients (sensitivities) of error functional \eqref{eq:J} with
respect to perturbations of, respectively, $D(c)$ and $t^+(c)$,
whereas $\xi_{D}^{(n)}$ and $\xi_{t^{+}}^{(n)}$ are the corresponding
lengths of the descent steps in the two directions. The initial guess
for problem P2 in \eqref{eq:descIG} is given by the constant values
$\widehat{D}$ and $\widehat{t^+}$ which are the optimal
reconstructions obtained from problem P1. (Problem P1 also requires an
initial guess, however, this problem tends to have a unique minimum
and therefore this initial guess may be arbitrary \cite{k14a}. {Thus,
  the advantage of solving problem P1 first is that it provides a very
  robust initial guess for problem P2}.)  The optimal
concentration-dependent properties can then be computed using
\eqref{eq:desc} as $\widehat{D}(c) = \lim_{n \rightarrow \infty}
D^{(n)}(c)$ and $\widehat{t^+}(c) = \lim_{n \rightarrow \infty}
t^{+(n)}(c)$. A key element of the iterative process \eqref{eq:desc}
is the evaluation of error functional gradients
$\nabla_{D}\mathcal{J}$ and $\nabla_{t^+}\mathcal{J}$ for which
details are provided in \ref{sec:grads}. We also refer the reader to
\cite{bvp11,bp11a} for a discussion of further mathematical and
computational details of this approach.

The estimates of the material properties obtained by solving problems
P1 and P2 are optimal, in the sense of minimizing the error with
respect to measurements, cf.~\eqref{eq:J}. Such inverse problems are
however known to be often ``ill-posed'', meaning that the presence of
noise in the measurement data may significantly affect the
reconstructed solution \cite{ehn96,t05}. The sensitivity of the
obtained reconstructions to perturbations of the data may be probed by
performing a Monte-Carlo analysis \cite{sethurajan2015accurate} in
which problems P1 and P2 are solved repeatedly using measurements
$\tilde{c}$ artificially contaminated with independent noise samples
with an assumed (e.g., normal) distribution and magnitude determined
by the known size of the measurement errors. While this approach
provides valuable insights about the sensitivity of the reconstructed
material properties to noise in the data, it does not quantify their
uncertainty in the sense of indicating which values of the material
properties are most likely. A solution to this problem is presented in
the next subsection.

\subsection{\sffamily \Large Bayesian Approach to Uncertainty Quantification}
\label{sec:bayes}

We assume here that both the measurements $\tilde{c}(x,t)$ and the
reconstructed material properties $[D,t^+]$, or $[D(c),t^+(c)]$, are
random variables characterized by certain probability density
functions (PDFs). More precisely, in the case of
concentration-dependent properties, $D(c)$ and $t^+(c)$ are given by
suitable probability distributions for {\em all} concentration values
$c \in [c_{\alpha}, c_{\beta}]$ and the same also applies to the
measurements $\tilde{c}$ for different values of $x \in [-0,L]$ and $t
\in [0,T]$.

In the Bayesian framework, the probability distribution of the
reconstructed material properties is given in terms of the {\em
  posterior} probability $\PP(m | \tc)$, which is the probability of
$m$ attaining a certain value (in problem P2, for a given concentration
$c$) given observations $\tc$, and can be expressed using Bayes' rule
\cite{s10,s13,t17}
\begin{equation}
  \PP(m|\tc)=\frac{\PP(\tc | m) \, \PP(m) }{\PP(\tc)}, 
\label{eq:bayes}
\end{equation}
where $\PP(m)$ is the {\em prior} distribution reflecting our a priori
assumptions about the solution, $\PP(\tc | m)$ is the {\em likelihood}
of observing particular experimental data for a given set of material
properties, whereas $\PP(\tc)$ is a normalizing factor.

In terms of the prior distribution $\PP(m)$, one can take the
distribution of $m$ obtained by performing a Monte-Carlo sensitivity
analysis of the deterministic inverse problems P1 and P2, as described
at the end of section \ref{sec:inverse}. This is accomplished by
solving problems P1 and P2 $N \ge 1$ times, each time using the
original measurements $\tc$ perturbed with normally-distributed noise
with magnitude given by the known size of experimental errors. The
obtained material properties $m$ are then used to construct the prior
distribution $\PP(m)$. This step appears as STAGE 1 in Algorithms
\ref{alg:P1} and \ref{alg:P2} below. We note that this analysis does
not account for how good the fits are, in terms of the value of the
error functional \eqref{eq:J}, for various samples of the noise
perturbing the measurements. A alternative, neutral, approach would be
to take ``uninformative'' priors given by uniform distributions of
$m$.

As regards the likelihood function, the following ansatz is typically
adopted in Bayesian inference \cite{s10,s13,t17}
\begin{equation}
  \PP(\tc | m) \propto e^{-\J(m)},
  \label{eq:likfun}
\end{equation}
which expresses the assumption that for a given set of material
properties $m$, measurements resulting in large values of the error
functional \eqref{eq:J} are less likely to be observed. The likelihood
function $\PP(\tc | m)$ is approximated by sampling the distribution
in \eqref{eq:likfun} using the Metropolis-Hastings algorithm
\cite{chib1995understanding} to produce $M$ samples. This algorithm is
based on the Markov-Chain Monte-Carlo (MCMC) approach
\cite{gilks1995markov} employed to randomize $m$ and at each step
involves solution of the governing system \eqref{eq:PN} for modified
(trial) material properties $m^*$ followed by the evaluation of the
error functional \eqref{eq:J}. At each step the algorithm moves in the
probability space collecting samples from the probability distribution
\eqref{eq:bayes}. A move in the probability space is accepted or
rejected based on a sample acceptance ratio $\gamma$ defined based on
the posterior distribution \eqref{eq:bayes} (see Algorithm
\ref{alg:P1} and \ref{alg:P2}). If one attempts to move to a point in
the probability space that is more probable than the existing point,
the move is accepted.  On the other hand, if one attempts to move to a
less probable point, the algorithm reject the move with some
probability based on the steepness of the probability decrease in the
given direction. Thus, the trajectory tends to sample frequently from
high-probability regions while occasionally also sampling from
low-probability regions.  The MCMC algorithm involves a ``burn-in''
process in which a certain number (usually the first $10\%$) of the
total number $M$ of accepted samples is discarded to avoid outliers
common at initial stages.

While application of the Metropolis-Hastings algorithm is fairly
straightforward in the finite-dimensional setting of problem P1, it is
more delicate in the continuous setting of problem P2. The main
difficulty is in constructing random perturbations of the
concentration-dependent material properties $D(c)$ and $t^+(c)$ in a
way that they will remain smooth enough for the Planck-Nernst system
\eqref{eq:PN} to be well defined (normally, these functions should be
at least once continuously differentiable and this issue is also
addressed at the end of \ref{sec:grads}). This is achieved by
parameterizing the material properties in terms of their truncated
cosine-series representations
\begin{align}
m_P(c) & = \frac{\hat{m}_0}{2}+\sum_{k=1}^{P} \hat{m}_k \cos\left[\frac{2\pi k (c - c_{\alpha})}{c_{\beta} - c_{\alpha}} \right], \quad c \in [c_{\alpha},c_{\beta}] \label{eq:cos} \\
\text{where} \quad \hat{m}_k &= \frac{2}{c_{\beta} - c_{\alpha}}
\int_{c_{\alpha}}^{c_{\beta}} m(c) \cos\left[\frac{2\pi k (c - c_{\alpha})}{c_{\beta} - c_{\alpha}} \right] \, dc, \quad k=1,\dots,P
\nonumber
\end{align}
and the number of terms $P$ is a discretization parameter. The choice
of the cosine-series expansion is dictated by the assumed behavior of
$D(c)$ and $t^+(c)$ at $c = c_{\alpha}, c_{\beta}$.  The
Metropolis-Hastings algorithm is initialized with a function $m_P(c)$
for which the cosine-series coefficients in \eqref{eq:cos} vanish as
$|\hat{m}_k| \sim \mathcal{O}(k^{-2})$. New trial samples are
generated by multiplying the cosine-series coefficients
$\hat{m}_1,\dots,\hat{m}_P$ by independent random numbers $\eta_k$,
$k=1,\dots,P$, chosen such that $|\eta_k| < C$ for all $k$, where $C >
0$ is a parameter. For sufficiently large $P$ this approach
approximates a continuous random distribution while preserving the
required smoothness of the trial material properties $[D(c),t^+(c)]$.
The proposed computational approach for Bayesian uncertainty
quantification is summarized as Algorithms \ref{alg:P1} and
\ref{alg:P2} for the problems with constant and
concentration-dependent material properties, respectively, and is
validated in the next section.

We note that, somewhat unconventionally, both the $prior$ distribution
and the $likelihood$ function are derived here from the same
experimental data, albeit in fundamentally different ways. In the
absence of other possibilities, an alternative solution would be to
use an uninformative prior. However, the proposed approach is
preferred as it results in tighter bounds.

\begin{algorithm}
\fontsize{10pt}{8pt}\selectfont{
\begin{algorithmic}
\STATE {\bf STAGE 1: Construct $N$ samples for prior distribution $\PP(m)$}
\REPEAT
\STATE perturb measurements $\tc$ with normally-distributed noise (magnitude given by the size of experimental errors)
\STATE find  $\widehat{m}$ by solving problem P1 (using function \texttt{fminsearch} and initial guess $m^{(0)}$)
\STATE store $\widehat{m}$ as a sample for prior distribution 
\UNTIL $N$ prior distribution samples are obtained
\STATE assimilate samples to construct $\mathbb P(\bar m)$
\STATE
\STATE {\bf STAGE 2: Construct $M$ samples for posterior distribution $\PP(m | \tc)$}
\STATE construct initial sample $\bar{m}^{(0)}$
\STATE $k \leftarrow 1$
\REPEAT
\STATE create a new trial position $\bar m^* = \bar m^k+\texttt{normrnd}(\bar 0 ,  C)$
\STATE calculate acceptance ratio $\gamma = \frac{\mathbb P(\bar m^{(*)}|\tc)}{\mathbb P( \bar m^{(k)}|\tc)}$ 
\STATE {\bf if} $\gamma \geq \texttt{rand}(1)$:  \quad $\bar m^{(k+1)}$=$\bar m^{(*)}$; \quad $k \leftarrow k+1$, 
\STATE {\bf else}: \qquad\qquad\quad \ discard $m^{(*)}$
\STATE $k=k+1$
\UNTIL $M+M/10$ samples are obtained for posterior distribution
\STATE discard the first $M/10$ samples 
\STATE assimilate the remaining samples to construct posterior probability distribution   $\PP(m | \tc)$
\end{algorithmic}
\caption{\fontsize{8pt}{5pt}\selectfont{: Two-stage algorithm to estimate the posterior
  probability distribution of constant material properties
  (\texttt{normrnd} and \texttt{fminsearch} are MATLAB
  functions).
  \newline \textbf{Input:} \\
  \hspace*{0.3cm} $\tc$ --- experimental data, \\
  \hspace*{0.3cm} $N,M$ --- numbers of samples generated in STAGE 1 and STAGE 2 \\
  \hspace*{0.3cm} $\varepsilon_{\J}$ --- tolerance in the solution of problem P1 in STAGE 1 \\
  \hspace*{0.3cm} $m^{(0)}$ --- initial guess in the solution of problem P1 in STAGE 1 \\
  \hspace*{0.3cm} $\bar{m}^{(0)}$ --- initial guess sample in STAGE 2 \\
  \hspace*{0.3cm} $C$ --- parameter controlling randomization in in STAGE 2 \\
  \textbf{Output:} \\
  \hspace*{0.3cm} an approximation of the posterior probability distribution $\PP(m|\tc)$}
}
\label{alg:P1}
}
\end{algorithm}

\begin{algorithm}
\fontsize{10pt}{8pt}\selectfont{
\begin{algorithmic}
\STATE {\bf STAGE 1: Construct N samples for prior distribution of $\mathbb P(\bar m)$}
\REPEAT
\STATE perturb measurements $\tc$ with normally-distributed noise (magnitude given by the size of experimental errors)
\STATE $\bar m^{(0)} \leftarrow \hat m$ (initial guess)
\STATE $n \leftarrow 1$
\REPEAT
\STATE solve governing system \eqref{eq:sys}
\STATE evaluate $\nabla_{\bar m} \mathcal J$.
\STATE compute the conjugate direction $\mathbf g\left[ \nabla_{\bar m} \mathcal J \right]$
\STATE perform line minimization: $\tau^{(n-1)}_{\bar m} $ $\leftarrow$ $\argminl_{\tau}\left \{\mathcal J \left ({\bar m}^{(n-1)}- \tau \, \mathbf g\left[\nabla_{\bar m }\mathcal J  \right ],\right ) \right \}$ 
\STATE update: $\bar m^{(n)} \ \leftarrow  \ \bar m^{(n-1)}- \tau^{(n-1)}_{\bar m} \, \mathbf g\left[\nabla_{\bar m} \mathcal J  \right ]$
\STATE $n \leftarrow n+1$
\UNTIL  $|\mathcal J(\bar m^{(n)})-\mathcal J(\bar m^{(n-1)})|<\,
\varepsilon_{\mathcal J}|\mathcal J(\bar m^{(n)}|$
\STATE store $\bar m^{(n)}$ as prior distribution sample
\UNTIL N prior distribution samples are obtained
\STATE assimilate samples to construct $\mathbb P(\bar m)$.
\STATE {\bf STAGE 2: Construct M samples for posterior distribution $  \mathbb P(\bar m|\tc)$}
\STATE $\bar m^{(0)} \leftarrow \bar m^i$
\STATE $k \leftarrow 1$
\REPEAT
\STATE create a new trial position $\bar f^* = \bar f^k \times \texttt{normrnd}(\bar 0 ,  C)$
\STATE using inverse Fourier transform obtain $\bar m^{(*)}$
\STATE calculate acceptance ratio $\gamma = \frac{\mathbb P(\bar m^{(*)}|\tc)}{\mathbb P( \bar m^{(k)}|\tc)}$ 
\STATE {\bf if} $\gamma \geq \texttt{rand}(1)$:  \quad $\bar m^{(k+1)}$=$\bar m^{(*)}$; \quad $k \leftarrow k+1$, 
\STATE {\bf else}: \qquad\qquad\quad \ discard $\bar m^{(*)}$ 
\UNTIL M+M/10 samples are obtained for posterior distribution
\STATE discard the first {$M/10$} samples
\STATE assimilate the remaining samples to obtain posterior probability distribution 
\end{algorithmic}
\caption{\fontsize{8pt}{5pt}\selectfont{: Two-stage algorithm to estimate the posterior
  probability distribution of constant material properties
  (\texttt{normrnd} and \texttt{fminsearch} are MATLAB
  functions).
  \newline \textbf{Input:} \\
  \hspace*{0.3cm} $\tc$ --- experimental data, \\
  \hspace*{0.3cm} $N,M$ --- numbers of samples generated in STAGE 1 and STAGE 2 \\
  \hspace*{0.3cm} $\varepsilon_{\J}$ --- tolerance in the solution of problem P2 in STAGE 1 \\
  \hspace*{0.3cm} $\hat m$ --- initial guess in the solution of problem P2 in STAGE 1, (here the solution of problem P1 is used as $\hat m$)\\
  \hspace*{0.3cm} $\bar{m}^i$ --- initial sample in STAGE 2 (chosen such that $\bar{m}^i \in \X$) \\
  \hspace*{0.3cm} $C$ --- parameter controlling randomization in in STAGE 2 \\
  \textbf{Output:} \\
  \hspace*{0.3cm} an approximation of the posterior probability distribution $\PP(m|\tc)$} }
\label{alg:P2}
}
\end{algorithm}

\section{\sffamily \Large RESULTS}


\subsection{Validation}
\label{sec:validation}

In this section we validate the Bayesian approach to uncertainty
quantification introduced in Section \ref{sec:bayes}. In addition to
establishing its consistency, this will also allow us to assess how
the results it produces depend on key numerical parameters and
properties of the data. We will do this for both problems P1 and P2
using an approach based on ``manufactured solutions''
\cite{bvp11,bp11a}, where certain values of $D$ and $t^+$ (in problem
P1), or functional forms of $D(c)$ and $t^+(c)$ (in problem P2), are
initially assumed and used to generate ''measurements'' by solving
system \eqref{eq:PN}. After being contaminated with noise of
prescribed magnitude, this data is used to solve inverse problems P1
and P2 and then to quantify the uncertainty of the obtained
reconstructions using Algorithms \ref{alg:P1} and \ref{alg:P2}. In
particular, this approach allows one to determine how the uncertainty
of the reconstructions depends on the level of noise in the data.

\begin{figure}
\centering
\begin{subfigure}[b]{0.475\textwidth}
\centering
\includegraphics[width=\textwidth]{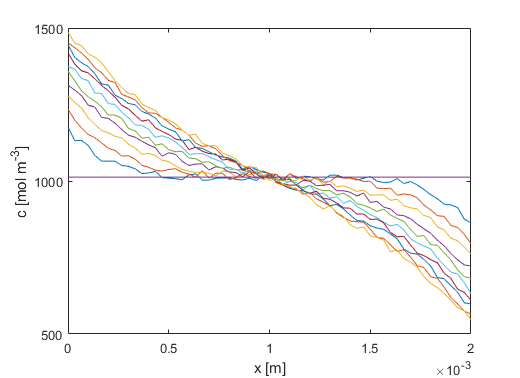}
\caption{}
\end{subfigure}
\hfill
\begin{subfigure}[b]{0.475\textwidth}
\centering
\includegraphics[width=\textwidth]{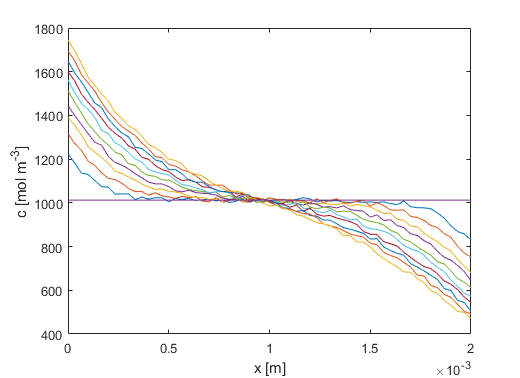}
\caption{}
\end{subfigure}
\caption{Concentration profiles $\tc(x,t_i)$, $i=1,\dots,N_T$,
  manufactured by solving problem \eqref{eq:PN} using (a) assumed
  constant material properties $[D,t^+]$ and (b) assumed
  concentration-dependent material properties $[D(c),t^+(c)]$. In both
  cases the added noise has variance $\xi = 25$ mol m$^{-3}$.}
\label{fig:data}
\end{figure}

\begin{figure}
\centering
\begin{subfigure}[b]{0.475\textwidth}
\centering
\includegraphics[width=\textwidth]{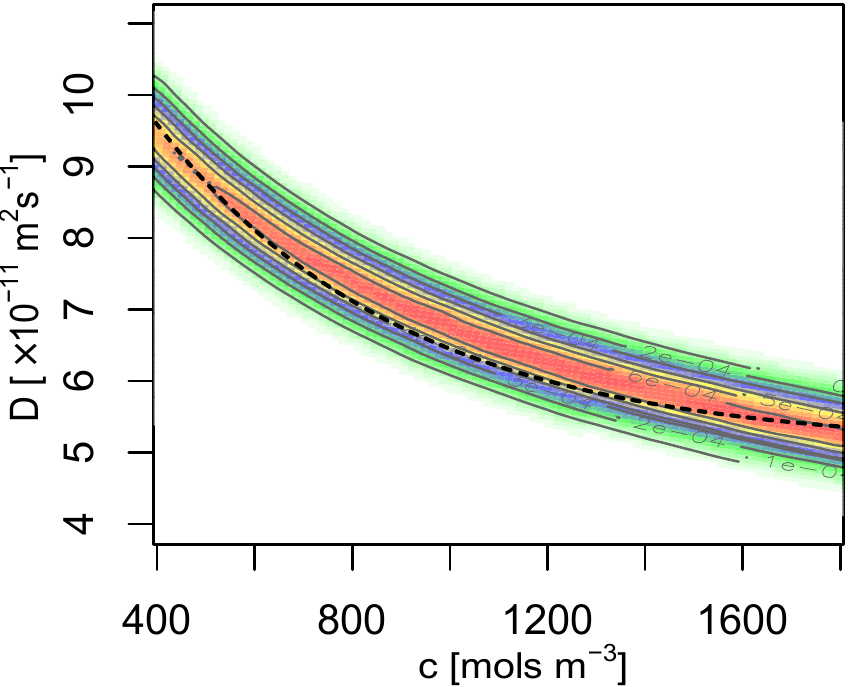}
\caption{ }
\label{fs:realD}
\end{subfigure}
\hfill
\begin{subfigure}[b]{0.475\textwidth}
\centering
\includegraphics[width=\textwidth]{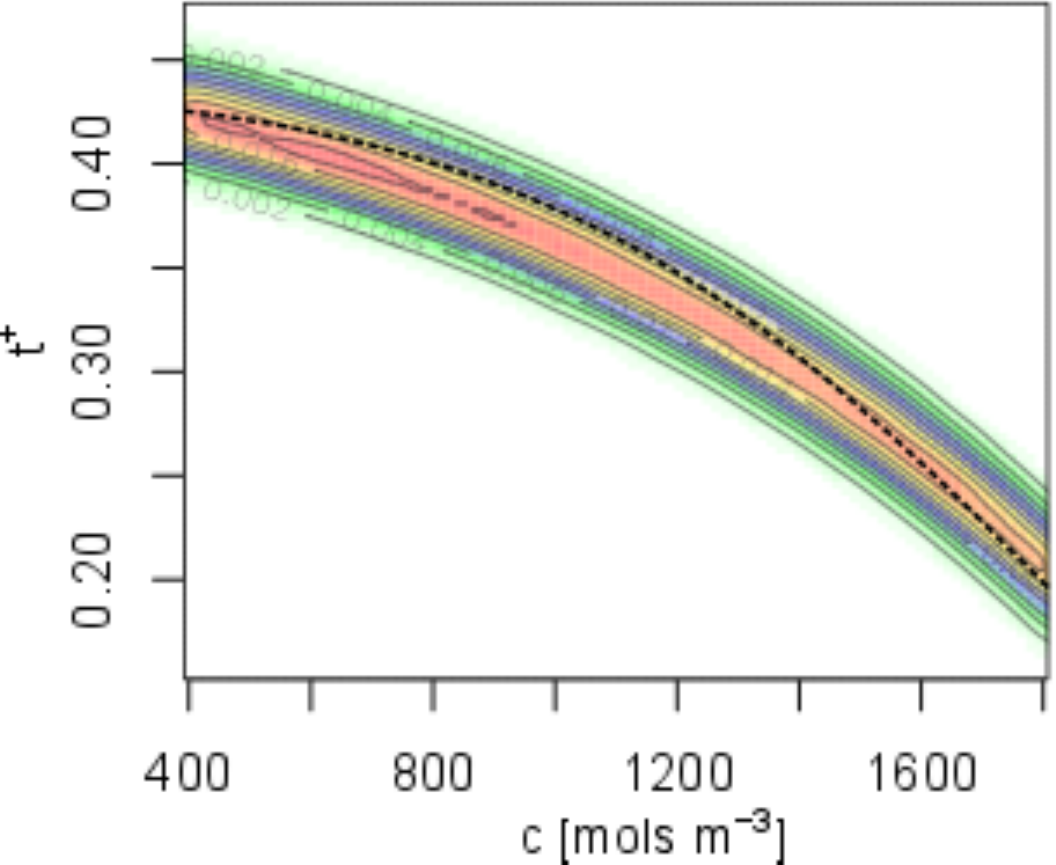}
\caption{ }
\label{fs:realtp}
\end{subfigure}
\caption{Posterior probability densities of (a) the diffusion
  coefficient $D(c)$ and (b) the transference number $t^+(c)$ as functions
  of the concentration $c$. Measurement data is available at $N_T =
  10$ time levels and the noise variance is $\xi = 25$ mol m$^{-3}$.
  The thick dashed lines represent the true distributions of $D(c)$
  and $t^+(c)$.}
\label{fig:validP2}
\end{figure}

For problem P1 we assume $D=10^{-10}$ m$^2$s$^{-1}$ and $t^+=0.4$,
whereas the assumed functional forms of $D(c)$ and $t^+(c)$ in problem
P2 are shown with thick dashed lines in Figures \ref{fig:validP2}a and
\ref{fig:validP2}b, respectively. We also assume that the
electrochemical cell has length $L = 0.002$ m and diameter $0.001$ m,
the applied current is $I = 100$ $\mu$A and the initial salt
concentration is $c^{\text{init}} = 1000$ mol m$^{-3}$, whereas the
duration of the experiment is $T = 20$ hours, all of which are in the
ballpark of parameters used in practice. System \eqref{eq:PN} and its
adjoint \eqref{eq:adjD} are solved numerically in MATLAB with the
routine {\tt pdepe} which uses adaptive spatial discretization and
adaptive time-stepping adjusted such that the relative and absolute
tolerances, respectively, $10^{-8}$ and $10^{-10}$ are satisfied at
all points in time and space.  Computed concentration profiles
recorded at $N_T = 10$ equispaced time levels $t_i$, $i=1,\dots,N_T$,
are used as the measurements $\tc(x,t_i)$ (the integral with respect
to time $t$ in \eqref{eq:J} is therefore replaced with summation over
$i=1,\dots,N_T$). Thus generated measurements are then perturbed with
normally-distributed noise with the frequency 20 kHz and variance $\xi
= 25$ mol m$^{-3}$. The concentration profiles $\tc(x,t_i)$,
$i=1,\dots,N_T$, obtained with constant $[D,t^+]$ and
concentration-dependent material properties $[D(c),t^+(c)]$ are shown
in Figures \ref{fig:data}a and \ref{fig:data}b, respectively. When
sampling the likelihood function $\PP(\tc | m)$ in Algorithm
\ref{alg:P2} expression \eqref{eq:cos} is used with $P = 50$ terms.

\begin{figure}
 \centering
\begin{subfigure}[b]{0.475\textwidth}
\centering
\includegraphics[width=\textwidth]{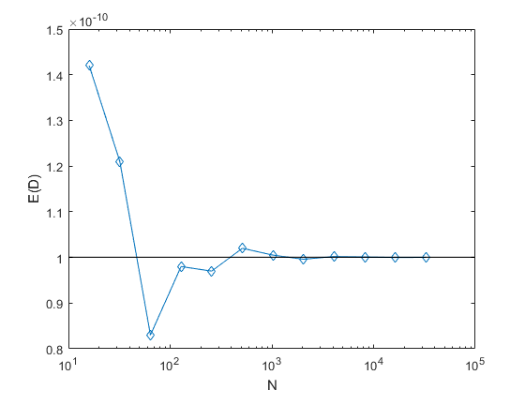}
\caption{}
\end{subfigure}
\hfill
\begin{subfigure}[b]{0.475\textwidth}
\centering
\includegraphics[width=\textwidth]{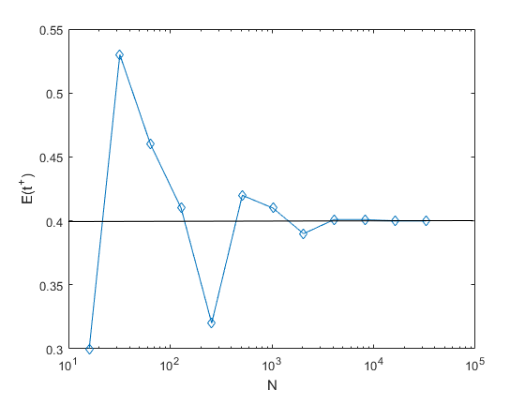}
\caption{}
\end{subfigure}
\caption{Convergence of the expected (constant) values of (a) $D$ and
  (b) $t^+$ to the true values indicated with horizontal lines as the
  number of samples $N$ and $M=N$ used in Algorithm \ref{alg:P1} is
  increased. Measurement data is available at $N_T = 10$ time levels
  and the noise variance is $\xi = 25$ mol m$^{-3}$.}
  \label{fig:conv}
\end{figure}

\begin{figure}
\centering
\includegraphics[width=0.475\textwidth]{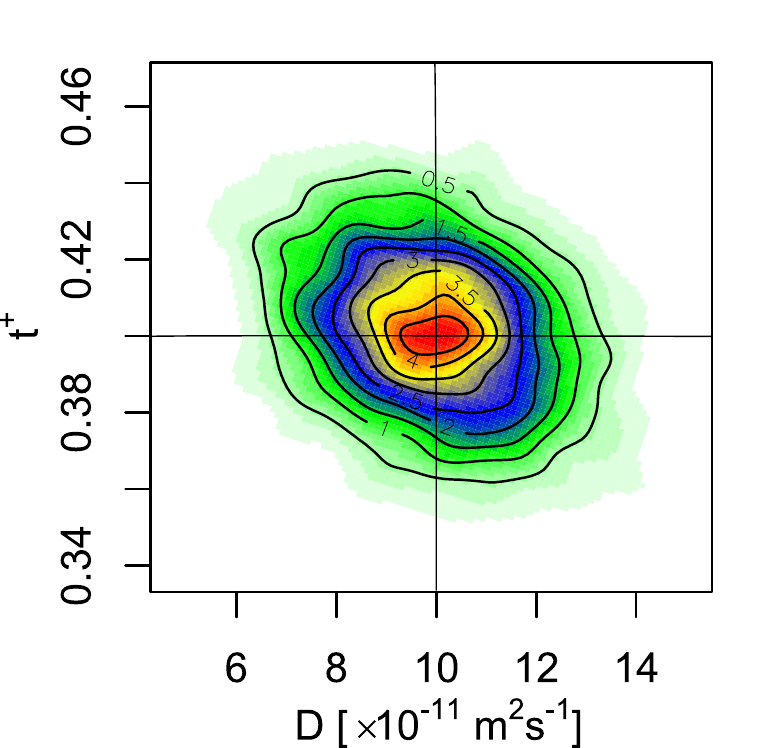}
\caption{Joint posterior probability distribution of the constant
  diffusion coefficient $D$ and transference number $t^+$. Measurement
  data is available at $N_T = 10$ time levels and the noise variance
  is $\xi = 25$ mol m$^{-3}$.}
\label{fig:validP1}
\end{figure}

We begin the presentation of the results by analyzing the effect of
the number of samples used to approximate the prior probability
distribution and the likelihood function $\PP(\tc | m)$ on the
convergence of the expected values of the constant material properties
in Figure \ref{fig:conv}. For simplicity, we set $M = N$ in Algorithm
\ref{alg:P1}.  In Figures \ref{fig:conv}a and \ref{fig:conv}b we see
that as the number of Monte-Carlo samples $N$ increases the expected
values of $D$ and $t^+$, estimated based on the posterior probability
distributions $\PP(m | \tc)$ produced by Algorithm \ref{alg:P1},
converge to their true values. Acceptable accuracy is achieved already
for $N = 5,000$ and this is the number of samples we will use below.
Next, in Figure \ref{fig:validP1}, we present the joint probability
density of the constant material properties $D$ and $t^+$ based on the
posterior distributions obtained with Algorithm \ref{alg:P1}. The
approximately symmetric shape of the isolines in this figure indicates
that there is no significant correlation between the uncertainties in
the reconstruction of the diffusion coefficient and the transference
number. The corresponding results obtained with Algorithm \ref{alg:P2}
for the problem with concentration-dependent material properties are
presented in Figures \ref{fig:validP2}a and \ref{fig:validP2}b for
$D(c)$ and $t^+(c)$, respectively, together with the corresponding
true distributions. The contour plots shown in these figures should be
interpreted such that their sections at a given value of $c$ produce
the posterior probability distributions functions $\PP(D(c) | \tc)$
and $\PP(t^+(c) | \tc)$. We observe that, unlike in Figure
\ref{fig:validP2}b where the most likely values of the transference
number $t^+(c)$ are quite close to the true distribution for all
values of $c$, in Figure \ref{fig:validP2}a a systematic difference
between the most likely reconstructed values of $D(c)$ and the true
values is evident. We remark here that in the absence of noise in the
data, the concentration-dependent diffusion coefficient $D(c)$
obtained by solving problem P2 is inferred very accurately and
coincides with the true distribution up to the graphical resolution
for all values of $c$ (this result is not shown). Hence, we can
conclude that the differences evident in Figure \ref{fig:validP2}a are
induced by noise and as such can be attributed to the ill-posedness of
the inverse problem (cf.~the discussion in Introduction).

\begin{figure}
\centering
\begin{subfigure}[b]{0.475\textwidth}
\centering
\includegraphics[width=\textwidth]{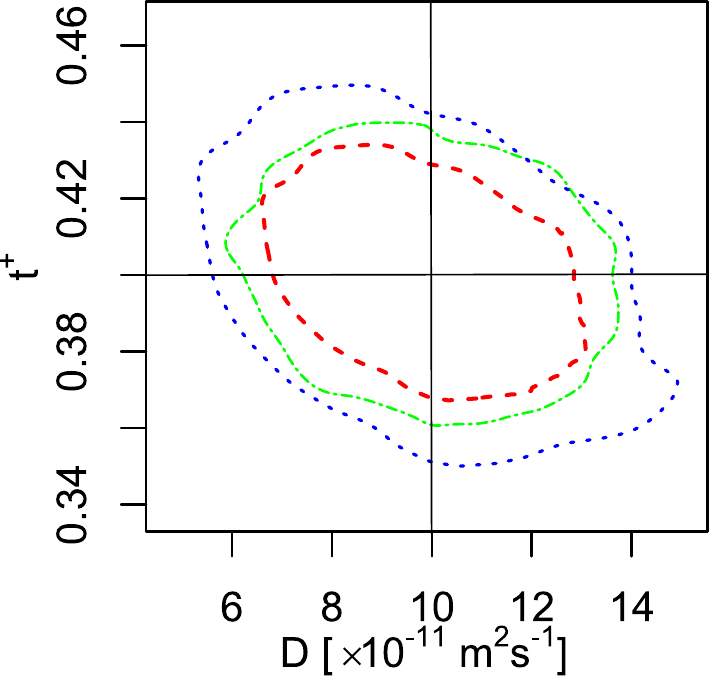}
\caption{ }
\label{fs:validcfb}
\end{subfigure}
\hfill
\begin{subfigure}[b]{0.475\textwidth}
\centering
\includegraphics[width=\textwidth]{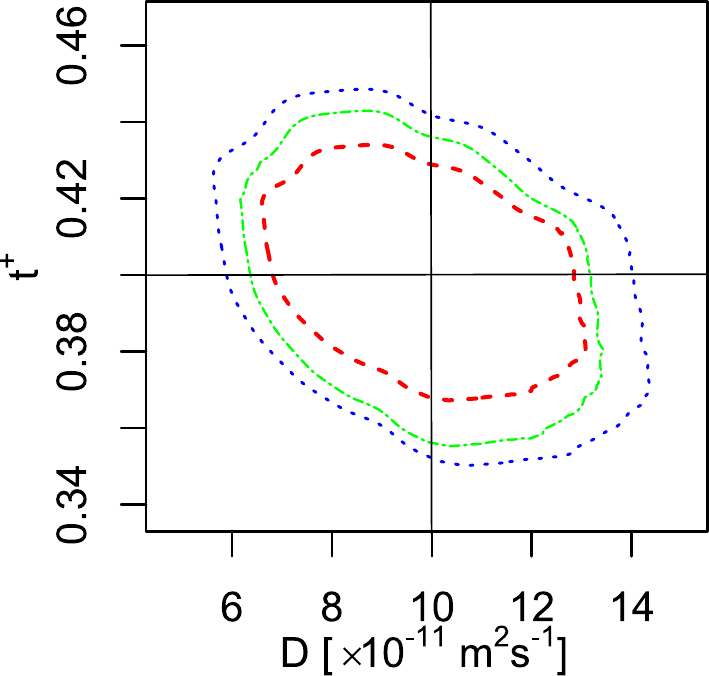}
\caption{ }
\label{fs:validcfb2}
\end{subfigure}
\caption{Boundaries of the $95\%$ confidence regions in the joint
  posterior probability distributions of the constant diffusion
  coefficient $D$ and transference number $t^+$ obtained with (a)
  concentration profiles $\tc(x,t_i)$ available at $N_T = 10$ time
  levels and perturbed with noise of different magnitudes (dashed ---
  $\xi = 25$ mol m$^{-3}$, dot-dash --- $\xi = 50$ mol m$^{-3}$,
  dotted --- $\xi = 75$ mol m$^{-3}$) , (b) concentration profiles
  $\tc(x,t_i)$ available at different numbers of time levels (dashed
  --- $N_T = 10$, dot-dash --- $N_T = 7$, dotted --- $N_T = 4$) and
  perturbed with noise of magnitude $\xi = 25$ mol m$^{-3}$.}
\label{fig:validP1nm}
\end{figure}

\begin{figure}
\centering
\begin{subfigure}[b]{0.475\textwidth}
\centering
\includegraphics[width=\textwidth]{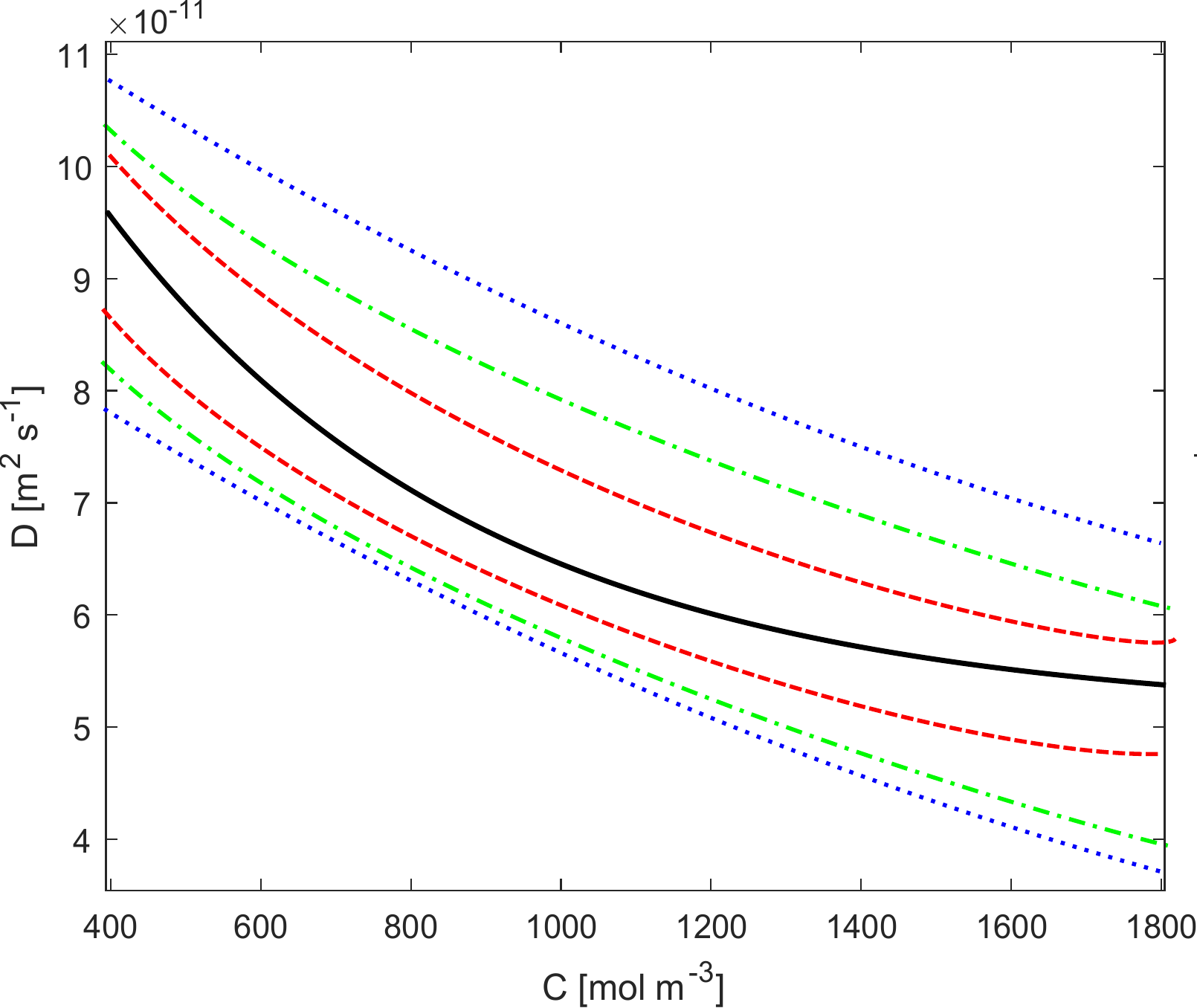}
\caption{ }
\label{fs:rb11}
\end{subfigure}
\hfill
\begin{subfigure}[b]{0.485\textwidth}
\centering
\includegraphics[width=\textwidth]{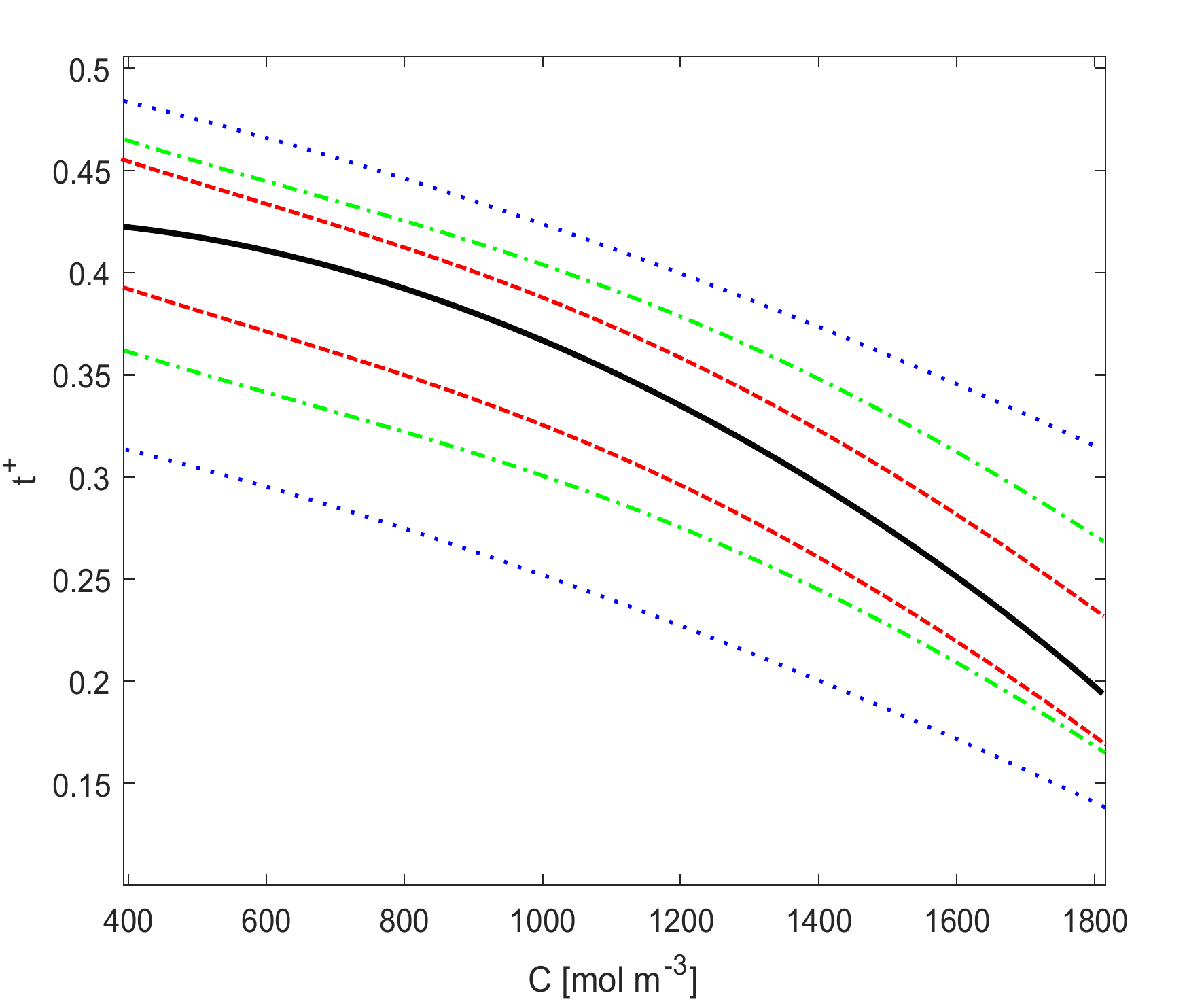}
\caption{ }
\label{fs:rb12}
\end{subfigure}
\caption{Boundaries of the $95\%$ confidence regions in the posterior
  probability distributions of (a) the diffusion coefficient $D(c)$
  and (b) the transference number $t^+(c)$ for different concentration
  values. The results are obtained using concentration profiles
  $\tc(x,t_i)$ available at $N_T = 10$ time levels and perturbed with
  noise of different magnitudes (dashed --- $\xi = 25$ mol m$^{-3}$,
  dot-dash --- $\xi = 50$ mol m$^{-3}$, dotted --- $\xi = 75$ mol
  m$^{-3}$). The thick solid lines represent the true distributions of
  $D(c)$ and $t^+(c)$.}
\label{fig:validP2n}
\end{figure}

We now move on to characterize the impact of the noise level in the
data $\tc$ on the uncertainty of the reconstructed material
properties. This is done by using noise with three different variances
$\xi = 25, 50, 75$ mol m$^{-3}$ and computing the posterior
distribution of the constant and concentration-dependent material
properties, $[D,t^+]$ and $[D(c),t^+(c)]$, using Algorithms
\ref{alg:P1} and \ref{alg:P2}. The results are presented,
respectively, in Figures \ref{fig:validP1nm}a and \ref{fig:validP2n},
where they are shown in terms of the $95 \%$ confidence bounds defined
as the boundaries of parameter regions over which the posterior
probability density integrates to $0.95$. We observe in these figures
that the confidence regions corresponding to different noise levels
have similar shapes and in all cases shrink as the noise level is
reduced, which is the expected behavior.

\begin{figure}
\centering
\begin{subfigure}[b]{0.475\textwidth}
\centering
\includegraphics[width=\textwidth]{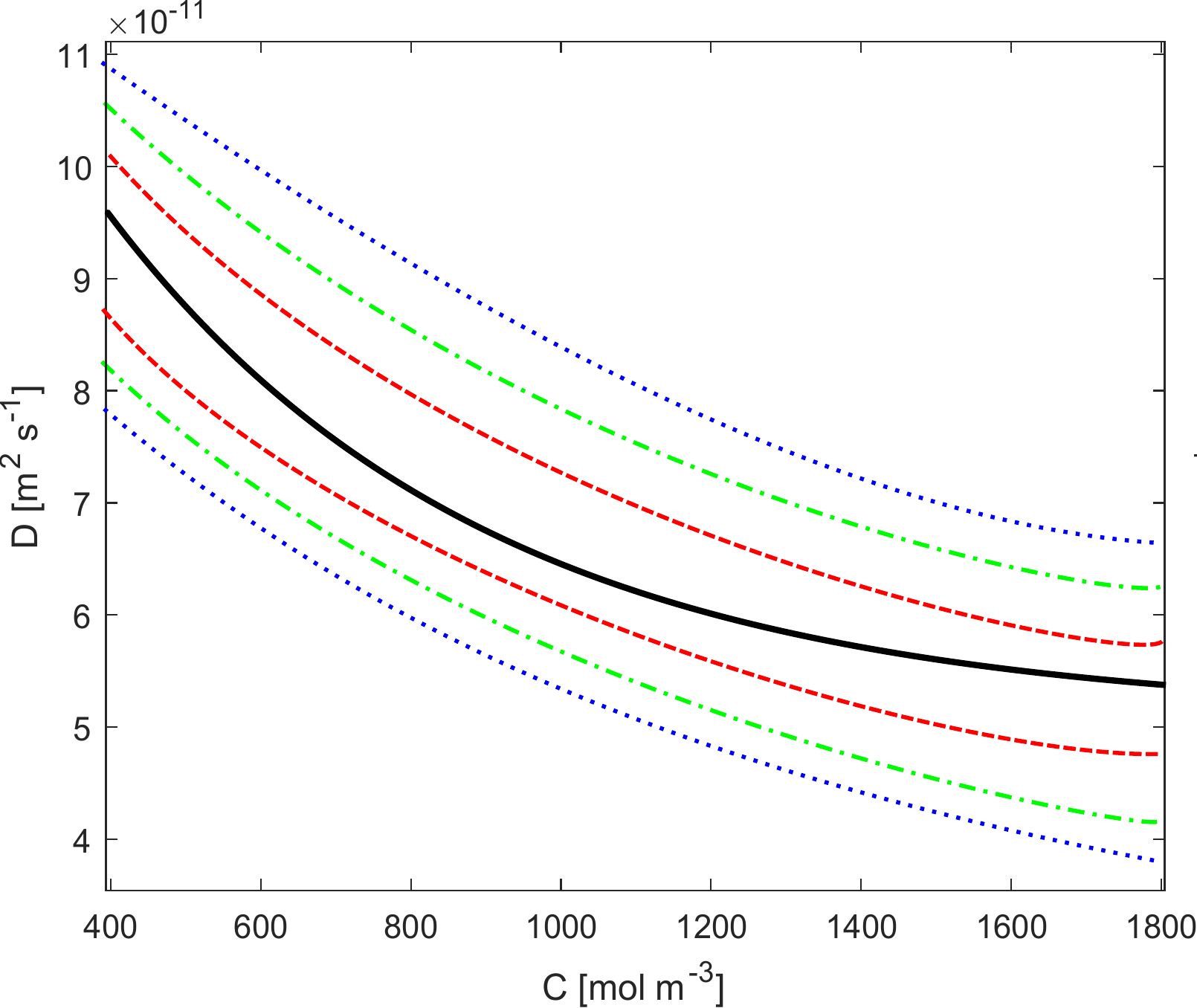}
\caption{ }
\label{fs:rb21}
\end{subfigure}
\hfill
\begin{subfigure}[b]{0.475\textwidth}
\centering
\includegraphics[width=\textwidth]{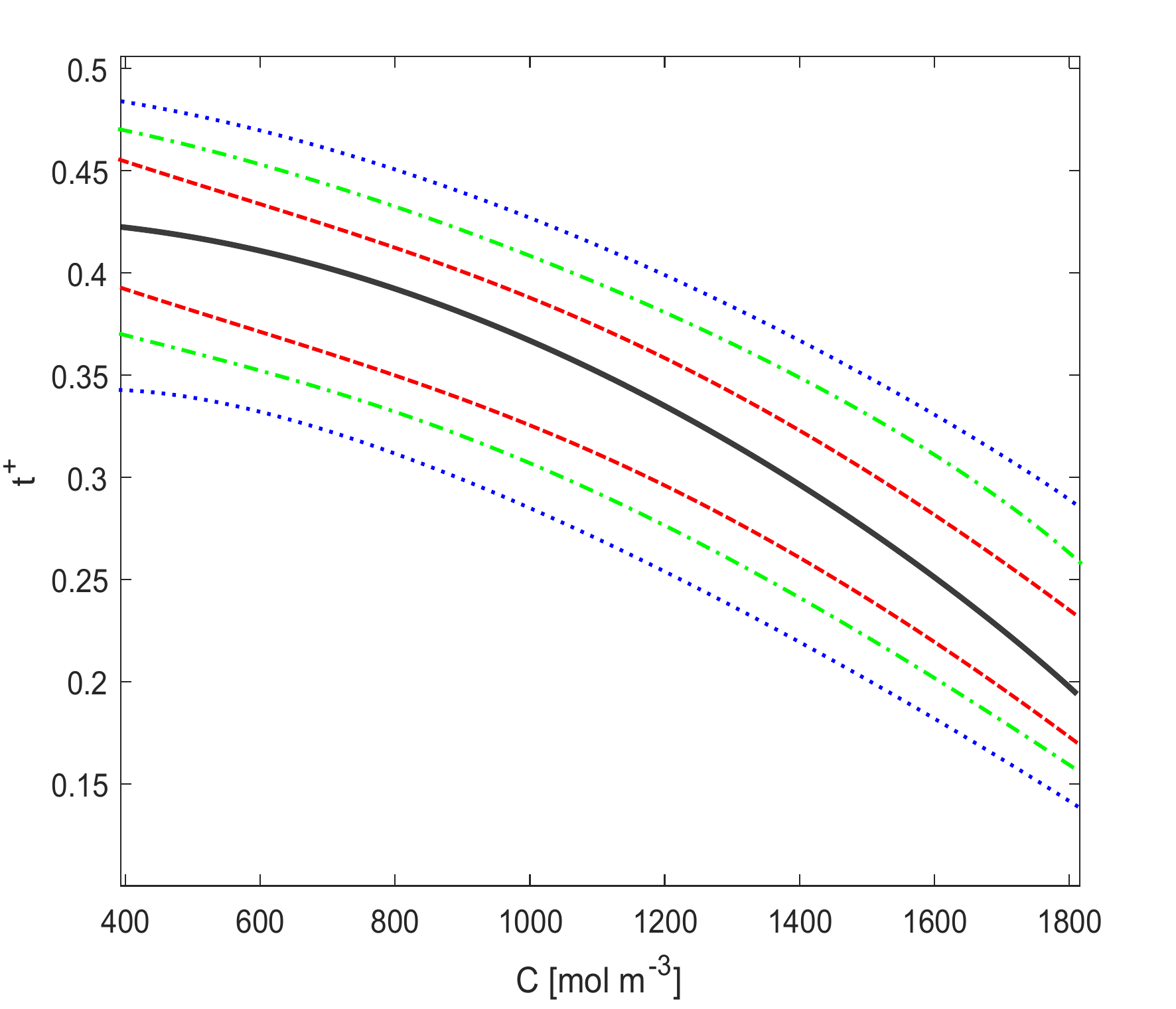}
\caption{ }
\label{fs:rb22}
\end{subfigure}
\caption{Boundaries of the $95\%$ confidence regions in the posterior
  probability distributions of (a) the diffusion coefficient $D(c)$
  and (b) the transference number $t^+(c)$ for different concentration
  values. The results are obtained using concentration profiles
  $\tc(x,t_i)$ available at different numbers of time levels (dashed
  --- $N_T = 10$, dot-dash --- $N_T = 7$, dotted --- $N_T = 4$) and
  perturbed with noise of magnitude $\xi = 25$ mol m$^{-3}$.  The
  thick solid lines represent the true distributions of $D(c)$ and
  $t^+(c)$.}
\label{fig:validP2m}
\end{figure}

We close this section by analyzing the effect of the amount of
available measurement data on the uncertainty of the reconstructed
material properties. This is done by varying the number $N_T$ of the
time levels $t_i$ where measurements $\tc(x,t_i)$, $i=1,\dots,N_T$,
are available ($N_T = 4, 7, 10$) while keeping the noise variance
fixed at $\xi = 25$ mol m$^{-3}$. The results obtained for problems
with constant and concentration-dependent material properties,
$[D,t^+]$ and $[D(c),t^+(c)]$, are presented, respectively, in Figures
\ref{fig:validP1nm}b and \ref{fig:validP2m}, again using the $95\%$
confidence bounds for the posterior probability distributions
determined with Algorithms \ref{alg:P1} and \ref{alg:P2}. We can
conclude from these figures that the effect of reducing the amount of
available data is qualitatively similar to the effect of increasing
the noise level in the data, as the uncertainty grows when $N_T$ is
decreased.

\subsection{\sffamily \Large Application to Experimental Data}
\label{sec:application}

\begin{figure}
\centering
\begin{subfigure}[b]{0.32\textwidth}
\centering
\includegraphics[width=\textwidth]{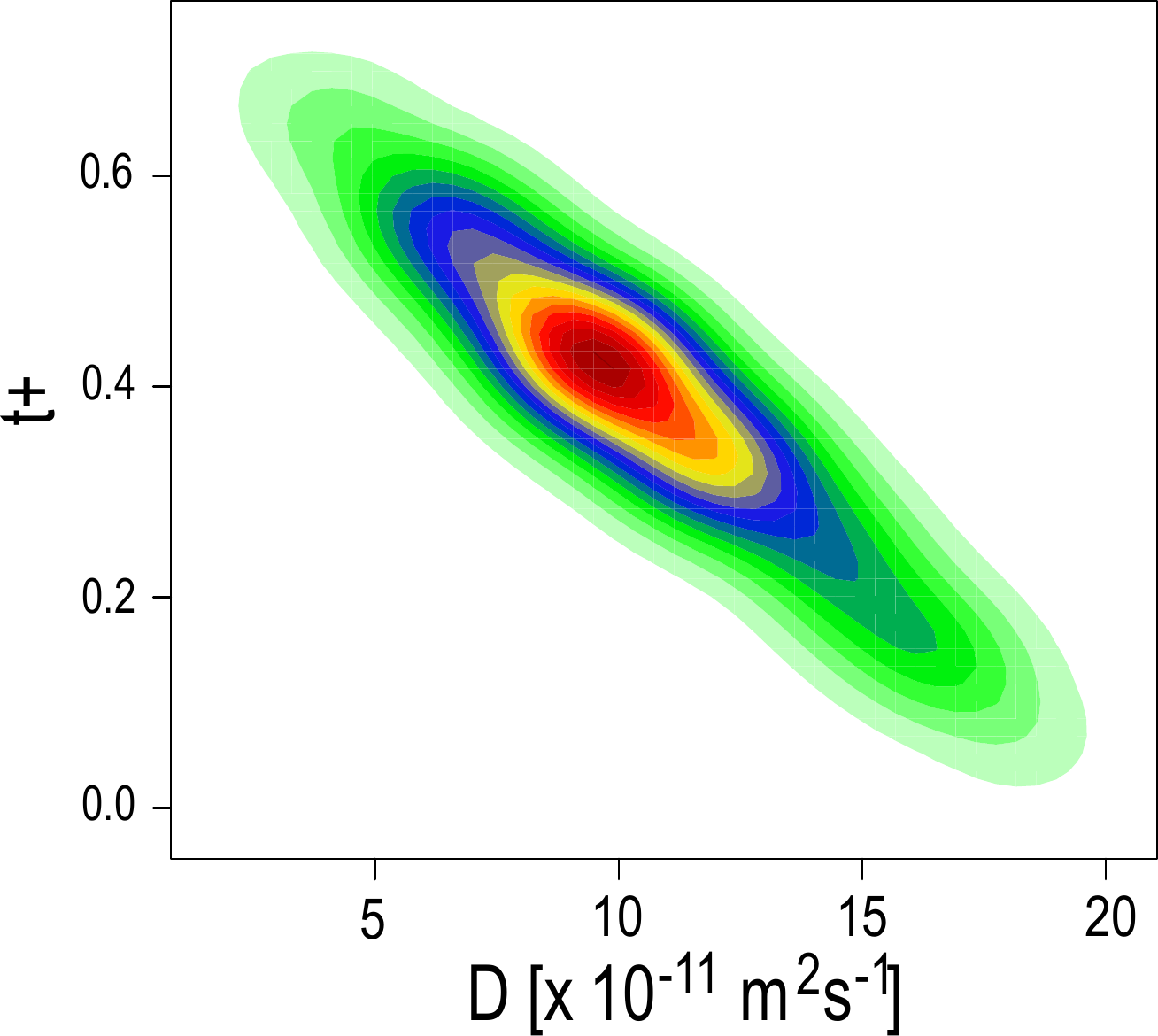}
\caption{ }
\label{fs:p1exp}
\end{subfigure}
\hfill
\begin{subfigure}[b]{0.32\textwidth}
\centering
\includegraphics[width=\textwidth]{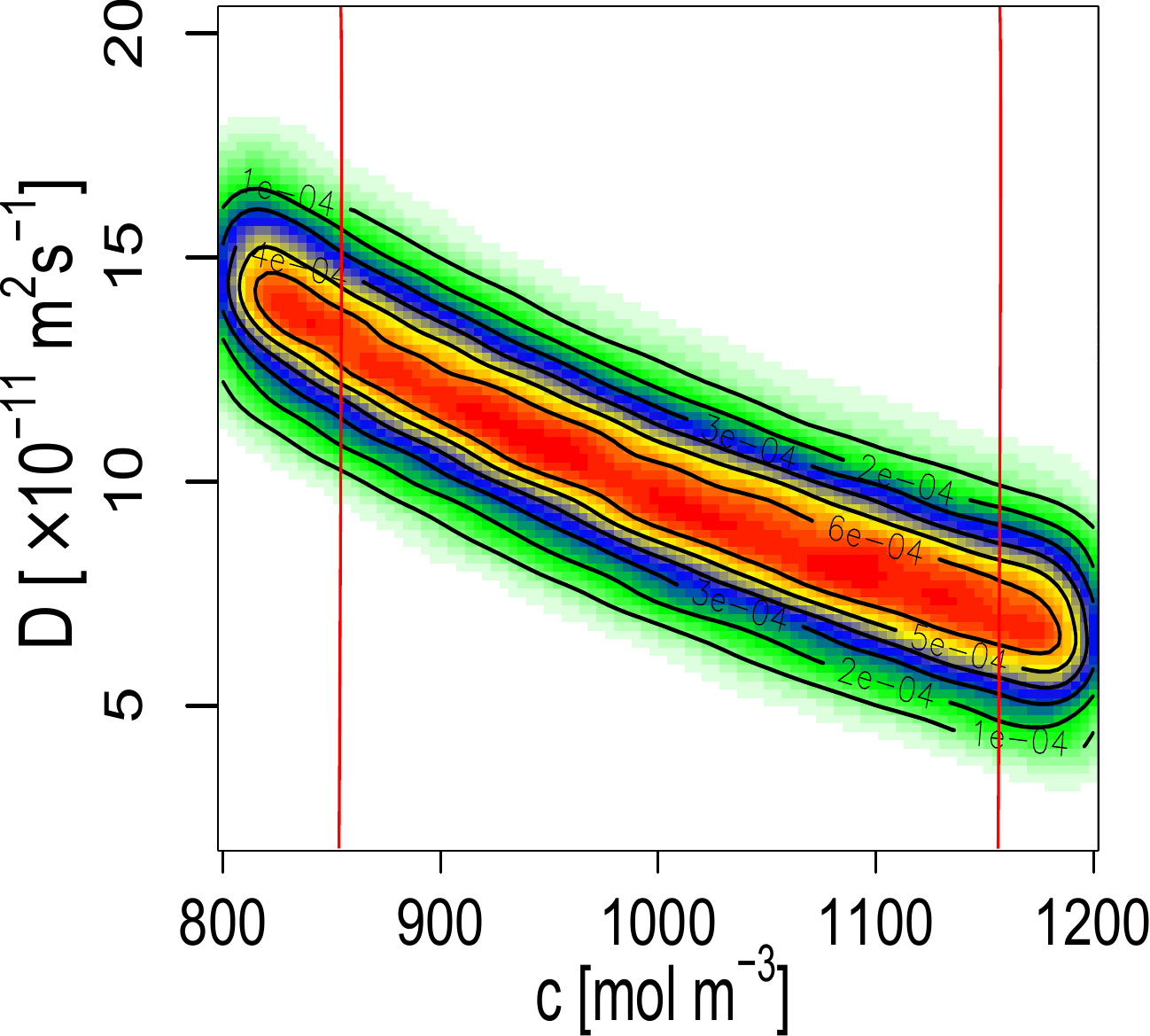}
\caption{ }
\label{fs:p2dexp}
\end{subfigure}
\hfill
\begin{subfigure}[b]{0.32\textwidth}
\centering
\includegraphics[width=\textwidth]{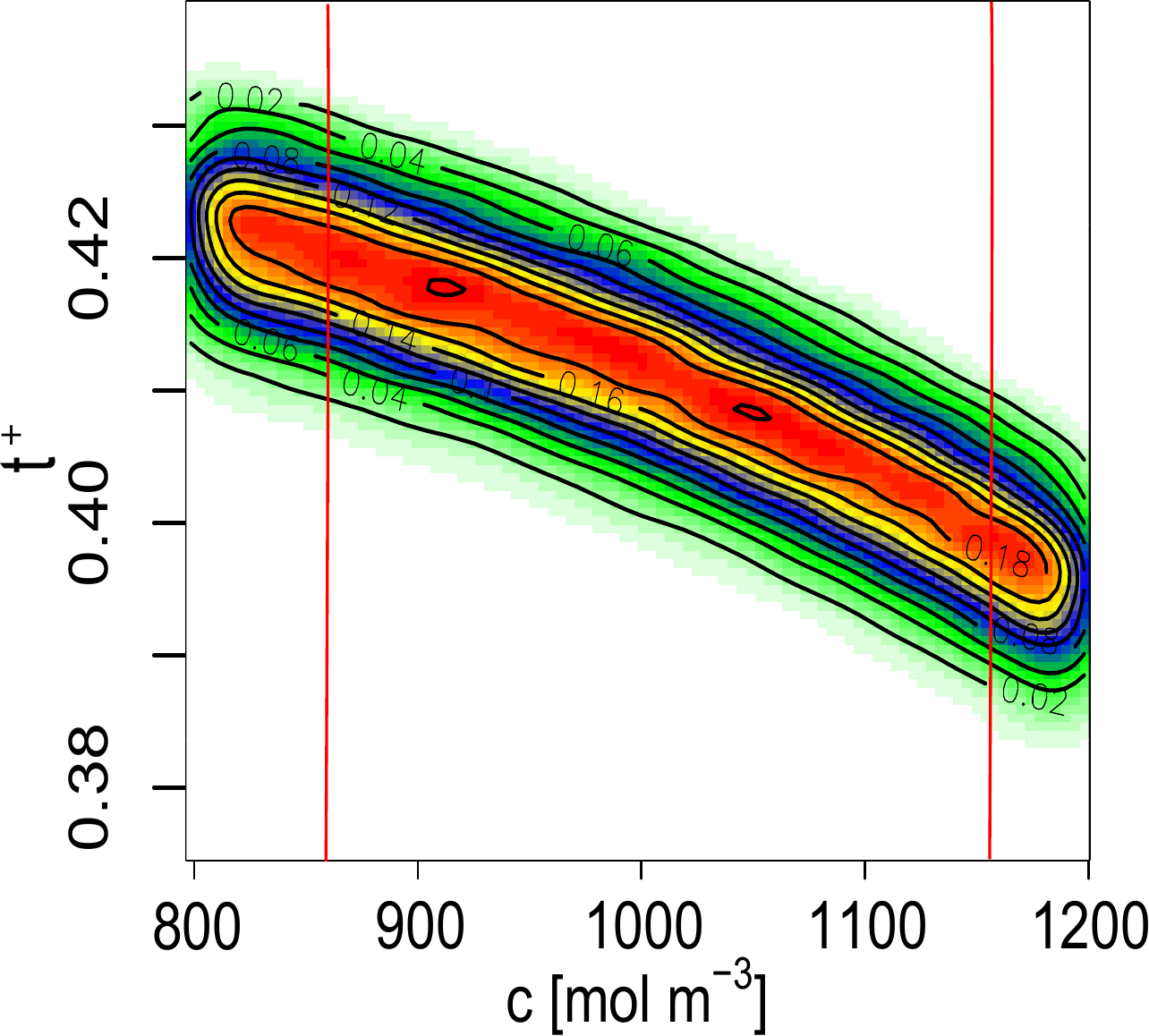}
\caption{ }
\label{fs:p2tpexp}
\end{subfigure}
\caption{(a) Joint posterior probability distribution of the constant
  diffusion coefficient $D$ and transference number $t^+$, posterior
  probability densities of (b) the diffusion coefficient $D(c)$ and
  (c) the transference number $t^+(c)$ as functions of the
  concentration $c$, all obtained based on the measurement data
  described in Section \ref{sec:experiment}.}
\label{fig:expP1P2}
\end{figure}

\begin{figure}
\centering
\begin{subfigure}[b]{0.32\textwidth}
\centering
\includegraphics[width=\textwidth]{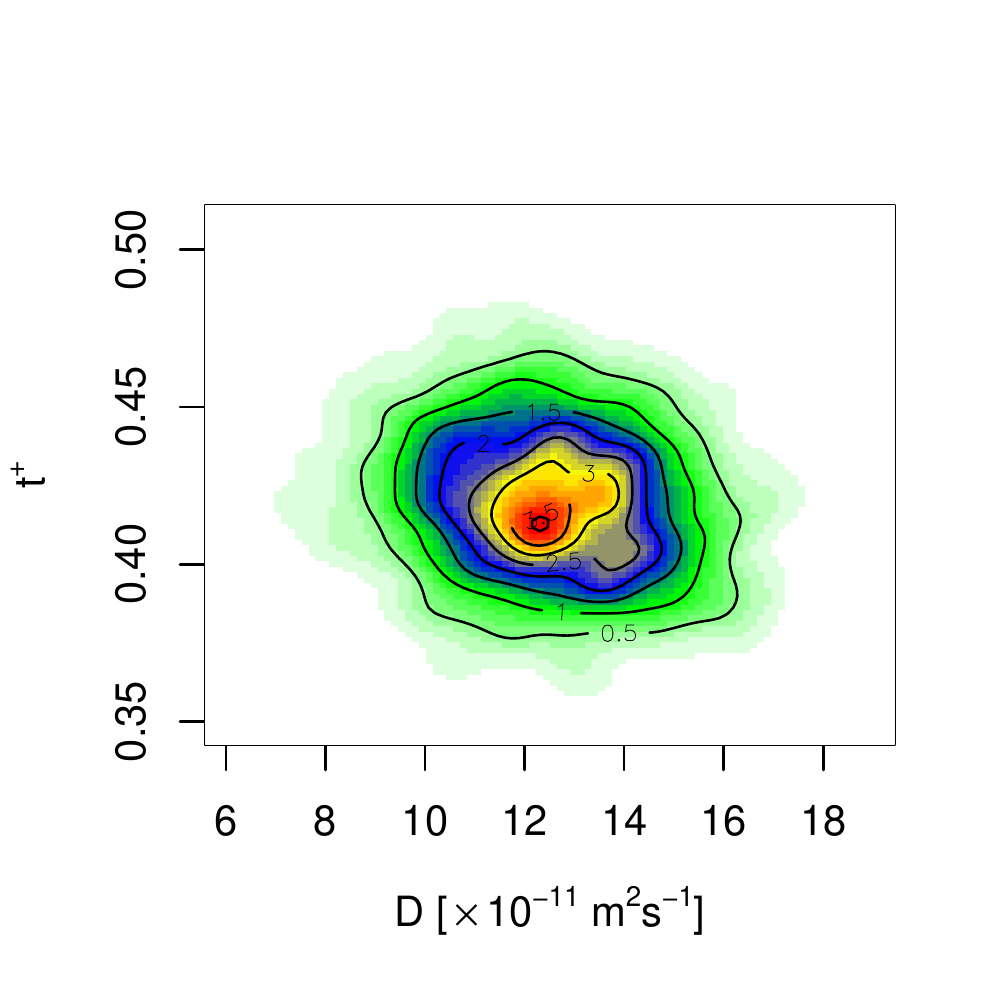}
\caption{ }
\label{fs:cp9m}
\end{subfigure}
\hfill
\begin{subfigure}[b]{0.32\textwidth}
\centering
\includegraphics[width=\textwidth]{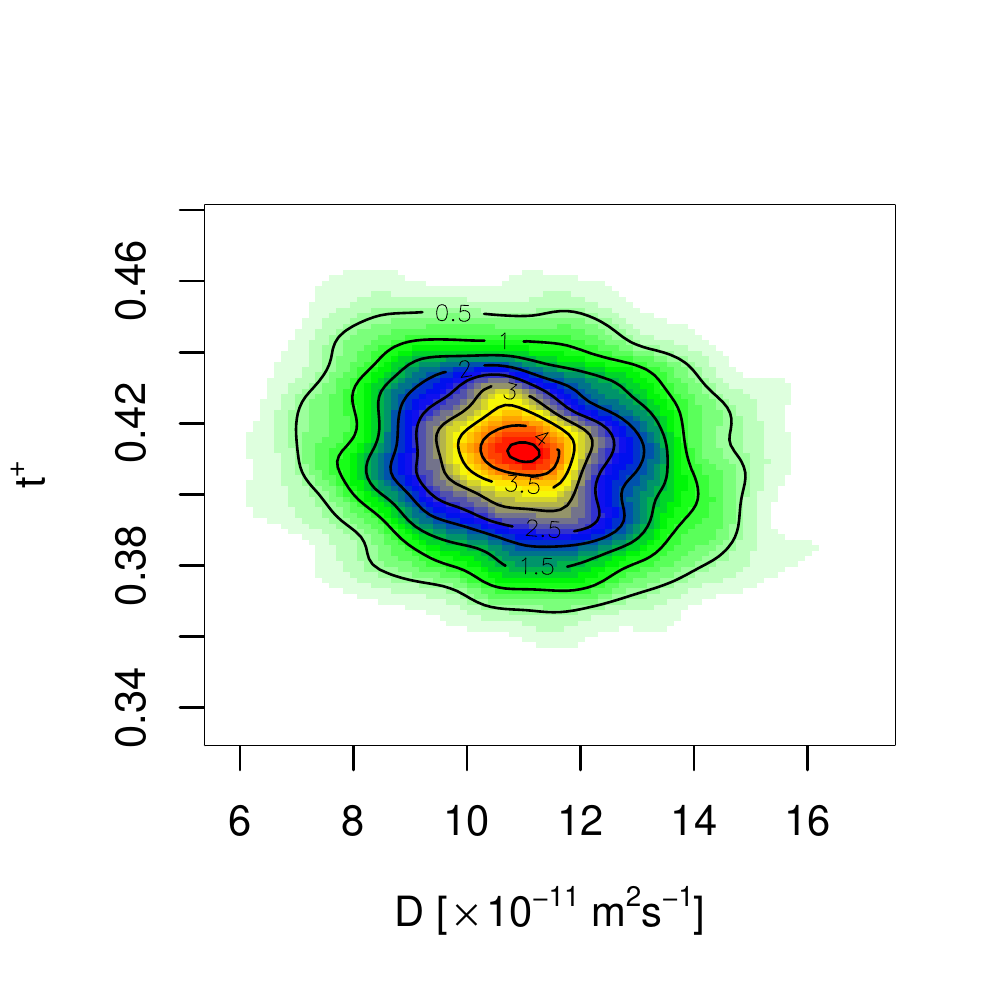}
\caption{ }
\label{fs:cp1m}
\end{subfigure}
\hfill
\begin{subfigure}[b]{0.32\textwidth}
\centering
\includegraphics[width=\textwidth]{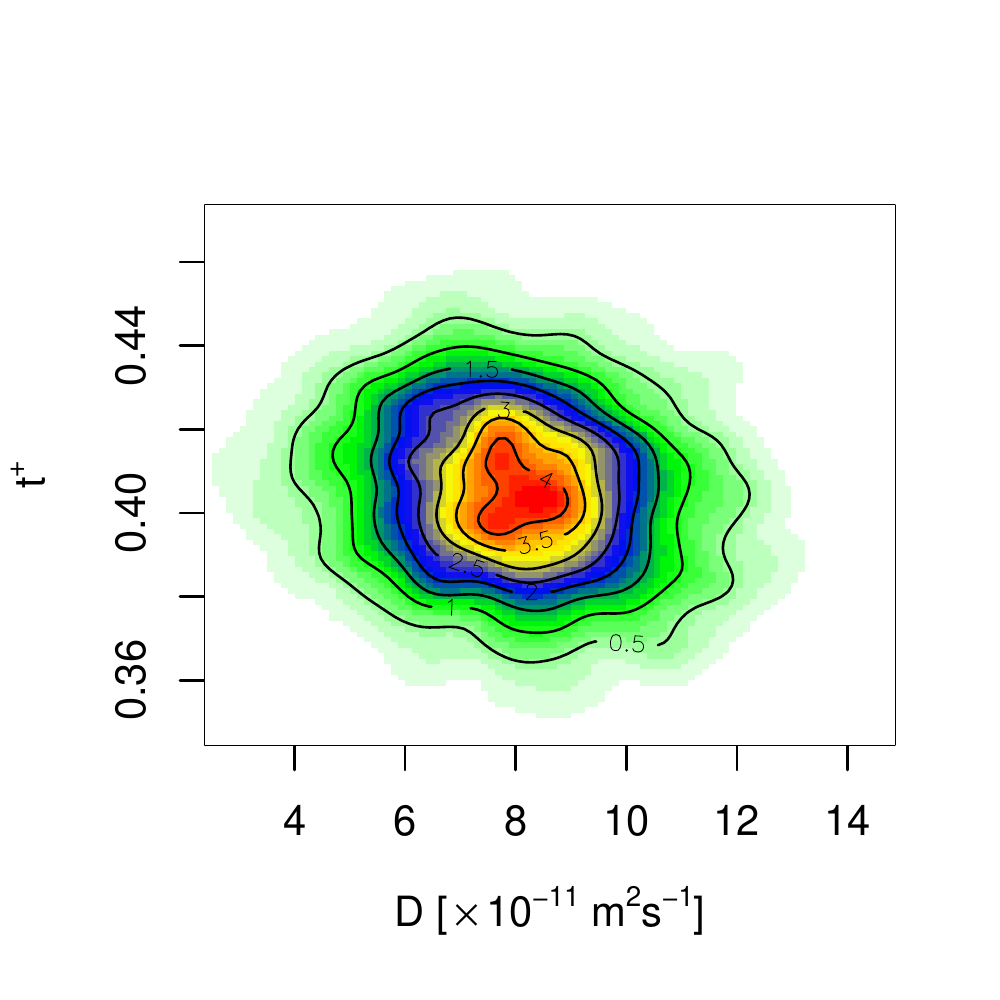}
\caption{ }
\label{fs:cp11m}
\end{subfigure}
\caption{Joint posterior probability distribution of the
  concentration-dependent diffusion coefficient $D(c)$ and
  transference number $t^+(c)$ at the concentrations (a) $c = 900$ mol
  m$^{-3}$, (b) $c = 1,000$ mol m$^{-3}$ and (c) $c = 1,100$ mol
  m$^{-3}$, cf.~Figures \ref{fig:expP1P2}b and \ref{fig:expP1P2}c,
  obtained based on the measurement data described in Section
  \ref{sec:experiment}.}
\label{fig:expP2c}
\end{figure}

In this section we apply the methodology for uncertainty
quantification described in Section \ref{sec:bayes} and validated in
Section \ref{sec:validation} to the inverse problems P1 and P2
involving, respectively, constant and concentration-dependent material
properties and using the experimental data described in Section
\ref{sec:experiment}. The joint posterior probability density obtained
for constant $[D,t^+]$ is shown in Figure \ref{fig:expP1P2}a, whereas
the posterior probability densities of $D(c)$ and $t^+(c)$ as
functions of the concentrations $c$ are shown in Figures
\ref{fig:expP1P2}b and \ref{fig:expP1P2}c. In addition, in Figures
\ref{fig:expP2c}a--\ref{fig:expP2c}c we also present the joint
posterior probability densities of $[D(c),t^+(c)]$ for three selected
concentration values (these distributions are extracted from the data
in Figures \ref{fig:expP1P2}b and \ref{fig:expP1P2}c by constructing
sections at the indicated values of $c$).

First, in Figure \ref{fig:expP1P2} we note that the expected values of
both constant and concentration-dependent material properties as well
as the trends with changes of the concentration revealed in the latter
case agree with the results known from the literature
\cite{sethurajan2015accurate}. In Figure \ref{fig:expP1P2}a we also
observe that the reconstructed constant material properties exhibit
significant uncertainties which, unlike the validation results from
Figure \ref{fig:validP1}, are correlated in the sense that larger
values of the diffusion coefficient $D$ are likely to occur together
with smaller values of the transference number $t^+$, and vice versa.
On the other hand, in the concentration-dependent case the
reconstruction uncertainty is significantly reduced for both $D(c)$
and $t^+(c)$ for all concentration values $c$. In both cases this
uncertainty is small relative to the variation of $D(c)$ and $t^+(c)$
over the entire range of $c$.  Moreover, Figures
\ref{fig:expP2c}a--\ref{fig:expP2c}c demonstrates that, in contrast to
the case of constant material properties, cf.~Figure
\ref{fig:expP1P2}a, in the concentration-dependent case there is no
significant correlation between the uncertainties of $D(c)$ and
$t^+(c)$ at particular values of $c$.

\section{\sffamily \Large CONCLUSIONS}
\label{sec:final}

In this study we have developed and carefully validated a
state-of-the-art Bayesian approach to quantify the uncertainty of
material properties reconstructed from experimental data. This
approach combines a recently developed inverse-modelling technique
capable of inferring general concentration-dependent material
properties subject to minimal assumptions
\cite{sethurajan2015accurate} with a Markov-Chain Monte-Carlo method
for sampling the likelihood function. We emphasize that while the
present study focuses on an electrochemical system modeled by the
Planck-Nernst equation \eqref{eq:PN}, the proposed approach is in fact
also applicable to a broad range of problems in chemistry where
macroscopic models are used. Extensive numerical tests of the method
confirm that it exhibits the expected behavior as different parameters
are varied.

Application of the proposed approach to actual experimental data
allows us to place rigorous ``error bounds'' in the reconstructed
material properties. These results demonstrate that while the
uncertainty can be non-negligible for constant material properties, it
is significantly reduced in the concentration-dependent case. The
reason for this is that the required regularity of $D(c)$ and $t^+(c)$
as functions of $c$ (cf.~the discussion at the end of \ref{sec:grads})
imposes some constraints on how rapidly the material properties can
vary with $c$. As a result, the reconstruction uncertainty is small
relative to the range of variation of both $D(c)$ and $t^+(c)$, which
offers confidence in the reliability of inverse modelling.

\subsection*{\sffamily \large ACKNOWLEDGMENTS}


The authors are thankful to J. M.~Foster and I.~Halalay for helpful
discussions. Funding for this research was provided by an NSERC
(Canada) Strategic Project Grant (\# STPGP 479258-15).


\clearpage

\appendix 
\section{Error Functional Gradients}
\label{sec:grads}

The gradients of the error functional \eqref{eq:J} with respect to
concentration-dependent properties $D(c)$ and $t^+(c)$ can be
calculated starting from the directional derivatives defined as
follows (to simplify the notation, we will not indicate the dependence
on $c$ in this appendix)
\begin{subequations}
\label{eq:der}
\begin{align} 
\mathcal J'([D,t^+];D')&=\lim_{\epsilon \to 0} \epsilon^{-1} \left[ \mathcal J([D+\epsilon D',t^+])-\mathcal J([D,t^+]) \right], \label{eq:dder} \\ 
\mathcal J'([D,t^+];t^{{+'}})&=\lim_{\epsilon \to 0} \epsilon^{-1} \left[ \mathcal J([D,t^++\epsilon t^{+ '}])-\mathcal J([D,t^+]) \right], \label{eq:tder}
\end{align}
\end{subequations}
where $D'$ and $t^{{+'}}$ are the perturbations of control variables
$D$ and $t^+$, respectively, and for simplicity of notation here and
below we omit the dependence on $c$. In order to identify expressions
for the gradients of the error functional as elements of a functional
(Hilbert) space, we use the Riesz representation theorem \cite{l69}
\begin{subequations}
\label{eq:rrep}
\begin{align} 
\mathcal J'([D,t^+];D') & = \Big\langle \nabla_{D} \mathcal J,D' \Big\rangle_{ \mathcal X},  \label{eq:rrep1} \\
\mathcal J'([D,t^+];t^{+'}) & = \Big\langle \nabla_{t^+} \mathcal J,t^{+'} \Big\rangle_{ \mathcal X}, \label{eq:rrep2}
\end{align}
\end{subequations}
where $\langle\cdot,\cdot\rangle_\X$ denotes the inner product in
functional space $\mathcal X$ (to be specified below). To fix
attention, we begin with the directional derivative \eqref{eq:dder}
with respect to the diffusion coefficient $D$ which can be evaluated as
follows
\begin{equation} 
\label{eq:pertc}
\mathcal{J}'([D,t^{+}];D')=
\sum_{i=1}^{N_T} \int_{0}^{T} \int_{0}^{L} \left[ c(x,t;[D,t^{+}])-\tilde{c}(x,t) \right]\delta(t-t_i)c'(x,t;D,D')\, dx \,dt,
\end{equation}
where $\delta(\cdot)$ is the Dirac delta distribution and $c'$ is the
solution of the PDE system obtained as a perturbation of the governing
system \eqref{eq:PN}. Then, the following transformations is invoked
\begin{equation} \label{eq:v}
V(x,t)=\int _{c_\alpha}^{c(x,t)} D(s)\, ds, \quad x \in [0,L] \ \text{and} \ t \in [0,T],
\end{equation}
where $c_\alpha = \min_{t\in[0,T], \; x\in[0,L]} c(x,t)$. We will
define the identifiability interval $\I = [c_\alpha,c_\beta]$, where
$c_\beta = \max_{t\in[0,T], \; x\in[0,L]} c(x,t)$, as the range of
concentration values spanned by solutions of \eqref{eq:PN}. To
simplify the notation, we also denote
\begin{equation}\label{eq:q}
Q(x,t) = \frac{(1-t^+)I}{FA}.
\end{equation}
Using these definitions, the perturbation system takes the form
\begin{subequations}
\label{eq:simsys}
\begin{alignat}{2} 
\frac{\partial c'}{\partial t}&=\frac{\partial}{\partial x}\left ( \frac{\partial V'}{\partial x}+Q' \right )
&\qquad &\text{in} \ (0,L) \times (0,T],  \label{eq:simsys2} \\
\left . \left (\frac{\partial V'}{\partial x}+Q' \right )\right |_{x=0, L}&=0 
&&\text{in} \ [0,T], \label{eq:simsys3} \\
 \left. c' \right |_{t=0}& = 0 
&&\text{in} \ [0,L],\label{eq:simsys4}
\end{alignat}
\end{subequations}
where the perturbation variables $V'$ and $Q'$ are expressed as
\begin{align} 
V'(x,t)&=\int _{c_\alpha}^{c(x,t)} D'(s)ds+D(c)c'(x,t;D'), \label{eq:V'} \\
Q'(x,t)&=-\left[ \left( V_m^s\right)\frac{dt^+}{dc}c'(x,t;D')+ \right] \frac{I}{FA}. \label{eq:Q'}
\end{align}
We now observe that directional derivative \eqref{eq:pertc} is not 
in a form consistent with Riesz representation \eqref{eq:rrep1},
because the perturbation variable $D'$ does not appear explicitly in
it, but is instead hidden (as $V'$, cf.~\eqref{eq:V'}) in the
perturbation system \eqref{eq:simsys}. In order to transform the
directional derivative \eqref{eq:pertc} into the Riesz form
\eqref{eq:rrep1} we will employ adjoint analysis.

Multiplying equation \eqref{eq:simsys2} by {\em adjoint} variable $c^*$
and integrating over the time and space domain, we get
\begin{equation} \label{eq:der1}
\int_0^L \int_0^T \frac{\partial c'}{\partial t}c^*\,dt\,dx=
\int_0^T \int_0^L\left [ \frac{\partial^2 V'}{\partial x^2}c^*+\frac{\partial Q'}{\partial x}c^* \right ]\,dx\,dt.
\end{equation}
\noindent By re-organizing equation \eqref{eq:der1} and integrating it
by parts with respect to both space and time we obtain
\begin{multline} \label{eq:der2}
\int_0^L \left\{[c'c^*]_0^T-\int_0^T \frac{\partial c^*}{\partial t}c'dt\right\} \,dx \\
=\int_0^T\left\{\left [\frac {\partial V'}{\partial x}c^*\right ]_0^L -\int_0^L \frac{\partial V'}{\partial x}\frac{\partial c^*}{\partial x}dx+[Q'c^*]_0^L- \int_0^L Q'\frac{\partial c^*}{\partial x}dx \right\}\,dt.
\end{multline}
Using equations \eqref{eq:simsys2}--\eqref{eq:simsys4} we can
eliminate a number of boundary terms after which we integrate the term
with $\frac{\partial V'}{\partial x}$ by parts one more time, so that
we arrive at
\begin{multline} \label{eq:der3}
\int_0^L \left\{[c'c^*]_{t=T}-\int_0^T \frac{\partial c^*}{\partial t}c'dt\right\}\,dx \\
=\int_0^T\left\{-\left [\frac {\partial c^*}{\partial x}V'\right ]_0^L +\int_0^L V'\frac{\partial^2 c^*}{\partial x^2}dx- \int_0^L Q'\frac{\partial c^*}{\partial x}dx \right\}\, dt.
\end{multline}
Now we assume that the adjoint system (defined on the same domain as
the governing system \eqref{eq:PN}) is in the form
\begin{subequations}
\label{eq:adjD}
\begin{align}
-\frac{\partial c^*}{\partial t}&=D\frac{\partial^2 c^*}{\partial x^2}+\frac{dt^+}{dc} \frac{I}{FA}\frac{\partial c^*}{\partial x}+
\sum_{i=1}^{N_T} \left (c-\tilde c \right )\delta(t-t_i) ,\\
\left.  \frac {\partial c^*}{\partial x} \right |_{x=0, L}&=0,\\
\left. c^* \right|_{t=T} &=0 \label{eq:adjDc}
\end{align}
\end{subequations}
which reduces identity \eqref{eq:der3} to the following expression for
the directional derivative of the error functional
\begin{equation}
\mathcal J'([D,t^+];D') = 
\int_0^T \int_0^L \left [\int _{c_\alpha}^{c(x,t)} D'(s)ds\right]\frac{\partial^2 c^*}{\partial x^2}\,dx\,dt.
\label{eq:dJ}
\end{equation}
We remark that adjoint system \eqref{eq:adjD} is in fact a {\em
  terminal} value problem, cf.~\eqref{eq:adjDc}, which means that it
needs to be integrated backwards in time (however, since the term with
the time derivative has a negative sign, the problem is well-posed).
Although this is not the function space we will ultimately use in the
computations, for now we set $\X = L^2(\I)$ meaning that our function
space consists of square-integrable functions of the concentration
$c$. The corresponding inner product, needed in \eqref{eq:rrep1}, is
\begin{equation}
\Big\langle \nabla_D^{L^2} \J, D' \Big\rangle_{L^2(\I)} = 
\int_{c_{\alpha}}^{c_{\beta}} \nabla_D\J(c) D'(c)\, dc.
\label{eq:ipL2}
\end{equation}
Changing the order of integration in \eqref{eq:dJ} and employing
\eqref{eq:ipL2} we arrive at the following expression for the $L^2$
gradient of the error functional
\begin {equation}\label{eq:gL2D}
\nabla_{D}^{L^2} \mathcal{J}(s)
=\int_{0}^{T} \int_{0}^{L} \chi_{\left [ c_{\alpha},c(x,t) \right ] }(s)\frac{\partial^2 c^*}{\partial x^2} \,dx \,dt, \qquad s \in [c_\alpha,c_\beta],
\end{equation}
where
$\chi_{\left [ a,b\right ] }=
\begin {cases}
1, \text{          }s  \in \left[a,b\right] \\
0,\text{          } s  \notin \left[a,b\right]
\end {cases}$.

Starting from the directional derivative \eqref{eq:tder} and
proceeding along the same lines as above we can derive an expression
for the $L^2$ gradient of the error functional with respect to the
concentration-dependent transference numbers
\begin{equation} 
\label{eq:gL2t}
\nabla_{t^{+}}^{L^2} \mathcal{J}(s)
=\int_{0}^{T} \int_{0}^{L}  \frac {\partial c^{*}}{\partial x} \delta(s-c(x,t)) \,dx \,dt, \qquad  s \in [c_\alpha,c_\beta],
\end{equation}
where as before $c^*$ is a solution of adjoint system \eqref{eq:adjD}.
Expressions \eqref{eq:gL2D} and \eqref{eq:gL2t} can be evaluated in a
straightforward manner using standard numerical techniques.

Above we derived gradient expressions in the $L^2$ space, however, as
pointed out in earlier studies \cite{bvp11,bp11a}, such gradients are
not suitable for the reconstruction of material properties, because
they can potentially be discontinuous and undefined outside
identifiability region $\mathcal I$. Therefore, in order to ensure
suitable smoothness and domain of definition of the gradients, we will
define them in the Sobolev space $H^1(\I)$ of functions of the
concentration $c$ with square-integrable derivatives, i.e., in problem
P2 we set $\X = H^1(\I)$. This space is characterized by the following
inner product, cf.~\eqref{eq:rrep1}--\eqref{eq:rrep2} (as we did
above, we focus here on $\nabla_{D} \mathcal{J}$ with the
transformation for $\nabla_{t^+} \mathcal{J}$ being analogous)
\begin{equation}
\Big\langle \nabla^{H^1}_D \mathcal J,D'\Big\rangle_{H^1(\I)}=
\int_{c_\alpha}^{c_\beta} \left ( \nabla^{H^1}_D \mathcal J\,D'+
\ell^2\frac{d  \nabla^{H^1}_D \mathcal J}{ds}\frac{d D'}{ds}\right)\, ds,
\label{eq:ipH1}
\end{equation}
where $\ell$ is a parameter with the meaning of a ``length-scale''.
Invoking again Riesz' representation theorem \cite{l69}, now for the
inner product \eqref{eq:ipH1} in the Sobolev space $H^1$, we obtain
from \eqref{eq:dder}
\begin{equation}\label{eq:sob1}
\mathcal J'([D,t^+];D')= \Big\langle \nabla ^{L^2}_D \mathcal J,D' \Big\rangle_{L^2}
=\Big\langle \nabla ^{H^1}_D \mathcal J,D'\Big\rangle_{H^1}.
\end{equation}
Using integration by parts we deduce from
\eqref{eq:ipH1}--\eqref{eq:sob1}
\begin{equation}
\int_{c_\alpha}^{c_\beta}\nabla ^{L^2}_D \mathcal J \, D'(s)\, ds
=\int_{c_\alpha}^{c_\beta}  \left ( \nabla^{H^1}_D \mathcal J\,D'-\ell^2\frac{d^2  \nabla^{H^1}_D \mathcal J}{ds^2} D' \right) \,ds + \left [\frac{d  \nabla^{H^1}_D \mathcal J}{ds} D' \right ]_{s=c_\alpha}^{s=c_\beta}
\end{equation}
and then, recognizing that the perturbation $D' \in H^1(\I)$ is
arbitrary except for the assumption that it satisfies the homogeneous
Neumann boundary conditions at the endpoints $c=c_\alpha,c_\beta$, we
arrive at the following second-order boundary value problem defining
the new smooth gradient $\nabla^{H^1}_D \mathcal J$ in terms of the
$L^2$ gradient obtained in \eqref{eq:gL2D}
\begin{subequations}
\label{eq:gradH1}
\begin{alignat}{2}
\nabla^{H^1}_D \mathcal J-\ell^2\frac{d^2  \nabla^{H^1}_D \mathcal J}{ds^2} &= \nabla ^{L^2}_D \mathcal J
& \qquad &\text{on} \ (c_\alpha,c_\beta), \\
\frac{d}{ds} \nabla^{H^1}_D \mathcal J & = 0 && c=c_\alpha,c_\beta. \label{eq:gradH1b}
\end{alignat}
\end{subequations}
Transformation of the $L^2$ gradient into $H^1$ Sobolev gradient can
be interpreted as a low-pass filtering which suppresses high-frequency
noise and this property is necessary to eliminate the discontinuities
which may potentially arise in the $L^2$ gradients \cite{pbh04}. The
degree of noise filtration is determined by the Sobolev parameter
$\ell$ with higher values of $\ell$ resulting in smoother Sobolev
gradients.  The boundary conditions \eqref{eq:gradH1b} imply a certain
behavior of the reconstructed material properties $\widehat{D}(c)$ and
$\widehat{t^+}(c)$ at the endpoints of the interval
$[c_\alpha,c_\beta]$, namely, that their derivatives with respect to
$c$ are unchanged as compared to the initial guesses $D^{(1)}$ and
$t^{+\, (1)}$, cf.~\eqref{eq:descIG}.  All reconstruction results
reported in this study have been obtained using Sobolev gradients in
minimization algorithm \eqref{eq:desc}.



\begin{thebibliography}{10}

\bibitem{broussely2004li}
M.~Broussely and G.~Archdale, ``Li-ion batteries and portable power source
  prospects for the next 5--10 years,'' {\em Journal of Power Sources},
  vol.~136, no.~2, pp.~386--394, 2004.

\bibitem{lu2013review}
L.~Lu, X.~Han, J.~Li, J.~Hua, and M.~Ouyang, ``A review on the key issues for
  lithium-ion battery management in electric vehicles,'' {\em Journal of Power
  Sources}, vol.~226, pp.~272--288, 2013.

\bibitem{nt04}
J.~Newman and K.~E. Thomas-Alyea, {\em {Electrochemical Systems}}.
\newblock John Wiley and Sons, 2004.

\bibitem{t05}
A.~Tarantola, {\em Inverse Problem Theory and Methods for Model Parameter
  Estimation}.
\newblock SIAM, 2005.

\bibitem{yu1999determination}
P.~Yu, B.~N. Popov, J.~A. Ritter, and R.~E. White, ``Determination of the
  lithium ion diffusion coefficient in graphite,'' {\em Journal of The
  Electrochemical Society}, vol.~146, no.~1, pp.~8--14, 1999.

\bibitem{prosini2002determination}
P.~P. Prosini, M.~Lisi, D.~Zane, and M.~Pasquali, ``{Determination of the
  chemical diffusion coefficient of lithium in LiFePO$_4$},'' {\em Solid State
  Ionics}, vol.~148, no.~1, pp.~45--51, 2002.

\bibitem{klett2012quantifying}
M.~Klett, M.~Giesecke, A.~Nyman, F.~Hallberg, R.~W. Lindström, G.~Lindbergh,
  and I.~Fur{\'o}, ``{Quantifying mass transport during polarization in a Li
  Ion battery electrolyte by in situ $^7$Li NMR imaging},'' {\em Journal of the
  American Chemical Society}, vol.~134, no.~36, pp.~14654--14657, 2012.

\bibitem{rahman2016electrochemical}
M.~A. Rahman, S.~Anwar, and A.~Izadian, ``Electrochemical model parameter
  identification of a lithium-ion battery using particle swarm optimization
  method,'' {\em Journal of Power Sources}, vol.~307, pp.~86--97, 2016.

\bibitem{ehn96}
H.~Engl, M.~Hanke, and A.~Neubauer, {\em Regularization of Inverse Problems}.
\newblock Dordrecht: Kluver, 1996.

\bibitem{v02}
C.~R. Vogel, {\em Computational Methods for Inverse Problems}.
\newblock SIAM, 2002.

\bibitem{sethurajan2015accurate}
A.~K. Sethurajan, S.~A. Krachkovskiy, I.~C. Halalay, G.~R. Goward, and
  B.~Protas, ``{Accurate characterization of ion transport properties in binary
  symmetric electrolytes using in situ NMR imaging and inverse modeling},''
  {\em The Journal of Physical Chemistry B}, vol.~119, no.~37,
  pp.~12238--12248, 2015.

\bibitem{s13}
R.~Smith, {\em Uncertainty Quantification: Theory, Implementation, and
  Applications}.
\newblock Computational Science and Engineering, SIAM, 2013.

\bibitem{t17}
L.~Tenorio, {\em An Introduction to Data Analysis and Uncertainty
  Quantification for Inverse Problems}.
\newblock Philadelphia, PA: Society for Industrial and Applied Mathematics,
  2017.

\bibitem{s10}
A.~M. Stuart, ``{Inverse problems: A Bayesian perspective},'' {\em Acta
  Numerica}, vol.~19, pp.~451--559, 2010.

\bibitem{kaipio2000statistical}
J.~P. Kaipio, V.~Kolehmainen, E.~Somersalo, and M.~Vauhkonen, ``Statistical
  inversion and monte carlo sampling methods in electrical impedance
  tomography,'' {\em Inverse Problems}, vol.~16, no.~5, p.~1487, 2000.

\bibitem{bergamaschi2000inverse}
P.~Bergamaschi, R.~Hein, M.~Heimann, and P.~J. Crutzen, ``{Inverse modeling of
  the global CO cycle: 1. Inversion of CO mixing ratios},'' {\em Journal of
  Geophysical Research: Atmospheres}, vol.~105, no.~D2, pp.~1909--1927, 2000.

\bibitem{bousquet1999inverse}
P.~Bousquet, P.~Peylin, P.~Ciais, M.~Ramonet, and P.~Monfray, ``{Inverse
  modeling of annual atmospheric CO$_2$ sources and sinks: 2. Sensitivity
  study},'' {\em Journal of Geophysical Research: Atmospheres}, vol.~104,
  no.~D21, pp.~26179--26193, 1999.

\bibitem{michalak2003method}
A.~M. Michalak and P.~K. Kitanidis, ``{A method for enforcing parameter
  nonnegativity in Bayesian inverse problems with an application to contaminant
  source identification},'' {\em Water Resources Research}, vol.~39, no.~2,
  2003.

\bibitem{rubin2010bayesian}
Y.~Rubin, X.~Chen, H.~Murakami, and M.~Hahn, ``{A Bayesian approach for inverse
  modeling, data assimilation, and conditional simulation of spatial random
  fields},'' {\em Water Resources Research}, vol.~46, no.~10, 2010.

\bibitem{saha2009prognostics}
B.~Saha, K.~Goebel, S.~Poll, and J.~Christophersen, ``{Prognostics methods for
  battery health monitoring using a Bayesian framework},'' {\em IEEE
  Transactions on instrumentation and measurement}, vol.~58, no.~2,
  pp.~291--296, 2009.

\bibitem{samadi2013online}
M.~F. Samadi, S.~M. Alavi, and M.~Saif, ``{Online state and parameter
  estimation of the Li-ion battery in a Bayesian framework},'' in {\em American
  Control Conference (ACC), 2013}, pp.~4693--4698, IEEE, 2013.

\bibitem{nbl08}
A.~Nyman, M.~Behm, and G.~Lindbergh, ``{Electrochemical Characterisation and
  Modelling of the Mass Transport Phenomena in $LiPF_6–EC–EMC$
  Electrolyte},'' {\em Electrochim. Acta}, vol.~53, no.~22, pp.~6356--6365,
  2008.

\bibitem{k14a}
A.~K. Sethurajan, ``Reconstruction of concentration-dependent material
  properties in electrochemical systems,'' Master's thesis, McMaster
  University, 2014.

\bibitem{bvp11}
V.~Bukshtynov, O.~Volkov, and B.~Protas, ``{On Optimal Reconstruction of
  Constitutive Relations},'' {\em Physica D: Nonlinear Phenomena}, vol.~240,
  no.~16, pp.~1228--1244, 2011.

\bibitem{bp11a}
V.~Bukshtynov and B.~Protas, ``Optimal reconstruction of material properties in
  complex multiphysics phenomena,'' {\em J. Comput. Phys.}, vol.~242,
  pp.~889--914, 2013.

\bibitem{chib1995understanding}
S.~Chib and E.~Greenberg, ``{Understanding the Metropolis-Hastings
  Algorithm},'' {\em The american statistician}, vol.~49, no.~4, pp.~327--335,
  1995.

\bibitem{gilks1995markov}
W.~R. Gilks, S.~Richardson, and D.~Spiegelhalter, {\em {Markov chain Monte
  Carlo in practice}}.
\newblock CRC press, 1995.

\bibitem{l69}
D.~Luenberger, {\em Optimization by Vector Space Methods}.
\newblock John Wiley and Sons, 1969.

\bibitem{pbh04}
B.~Protas, T.~R. Bewley, and G.~Hagen, ``{A Computational Framework for the
  Regularization of Adjoint Analysis in Multiscale PDE Systems },'' {\em J.
  Comput. Phys.}, vol.~195, no.~1, pp.~49--89, 2004.

\end{thebibliography}

\end{document}